\documentclass[a4paper, amsfonts, amssymb, amsmath, reprint, showkeys, nofootinbib, twoside,superscriptaddress]{revtex4-2}
\usepackage[english]{babel}
\usepackage[utf8]{inputenc}
\usepackage[colorinlistoftodos, color=green!40, prependcaption]{todonotes}
\usepackage[left=15mm,right=15mm,top=20mm,columnsep=15pt]{geometry} 
\usepackage[pdftex, pdftitle={Article}, pdfauthor={Author}]{hyperref} 
\begin{document}
\title{Inverse Problem Approach to Aberration Correction for \textit{in vivo} Transcranial Imaging Based on a Sparse Representation of Contrast-enhanced Ultrasound Data}

\author{Paul Xing}
\affiliation{Department of Engineering Physics, Polytechnique Montréal, Montreal, Canada}
\author{Antoine Malescot}
\affiliation{Department of Physiology and Pharmacology, Université de Montréal, Montreal, Canada}
\affiliation{Department of Stomatology, Université de Montréal, Université de Montréal, Montreal, Canada}
\author{Eric Martineau}
\affiliation{Department of Physiology and Pharmacology, Université de Montréal, Montreal, Canada}
\affiliation{Department of Stomatology, Université de Montréal, Université de Montréal, Montreal, Canada}
\author{Ravi L. Rungta}
\affiliation{Department of Stomatology, Université de Montréal, Université de Montréal, Montreal, Canada}
\affiliation{Department of Neuroscience, Université de Montréal, Montreal, Canada}
\affiliation{Centre interdisciplinaire de recherche sur le cerveau et l’apprentissage (CIRCA), Université de Montréal, Montreal, Canada}
\author{Jean Provost}
\email[Correspondence email address: ]{jean.provost@polymtl.ca}
\affiliation{Department of Engineering Physics, Polytechnique Montréal, Montreal, Canada}
\affiliation{Montreal Heart Institute, Montreal, Canada}

\date{\today}

\begin{abstract}
Transcranial ultrasound imaging is currently limited by attenuation and aberration induced by the skull. First used in contrast-enhanced ultrasound (CEUS), highly echoic microbubbles allowed for the development of novel imaging modalities such as ultrasound localization microscopy (ULM). Herein, we develop an inverse problem approach to aberration correction (IPAC) that leverages the sparsity of microbubble signals. We propose to use the \textit{a priori} knowledge of the medium based upon microbubble localization and wave propagation to build a forward model to link the measured signals directly to the aberration function. A standard least-squares inversion is then used to retrieve the aberration function. We first validated IPAC on simulated data of a vascular network using plane wave as well as divergent wave emissions. We then evaluated the reproducibility of IPAC \textit{in vivo} in 5 mouse brains. We showed that aberration correction improved the contrast of CEUS images by 4.6 dB. For ULM images, IPAC yielded sharper vessels, reduced vessel duplications, and improved the resolution from 21.1 $\mu$m to 18.3 $\mu$m. Aberration correction also improved hemodynamic quantification for velocity magnitude and flow direction.
\end{abstract}


\maketitle

\section{Introduction}

Non-invasive imaging of the brain vasculature is of great clinical interest since alterations of the cerebral blood flow are present in multiple pathologies including stroke and neurodegenerative conditions such as Alzheimer's disease\cite{iadecola2017neurovascular}. By using the transmission of plane waves\cite{montaldo2009coherent} and moving the image formation to the software end, ultrafast ultrasound imaging increased by more than a 100-fold the frame rate of standard ultrasound imaging techniques while preserving a large field of view\cite{tanter2014ultrafast}. Such increase in frame rate led to the development of a multitude of novel applications to study hemodynamic changes in the brain vasculature, such as ultrafast Doppler and functional ultrasound imaging\cite{mace2011functional}. Furthermore, the development of ultrasound localization microscopy (ULM) \cite{christensen2014vivo, errico2015ultrafast} and later dynamic ULM \cite{bourquin2021vivo} allowed to measure hemodynamic changes such as pulsatility at the microscopic scale.

However, image formation algorithms used in ultrafast ultrasound can suffer from some inherent limitations. The image is most often reconstructed in post-processing by calculating the time of flight associated to each point of the image in a process called delay-and-sum (DAS) beamforming \cite{perrot2021so}. A constant speed-of-sound hypothesis is generally assumed but is not always valid for \textit{in vivo} imaging. The mismatch between real transit times and the computed time-of-flight causes what is referred to as phase aberration, characterized by the image degradation and a decrease in both contrast and resolution \cite{anderson2000impact, pinton2011sources} which can represent an important limitation, especially for transcranial brain imaging. In ULM, skull-induced aberration can lead to detection errors\cite{mccall2021characterization}, which can cause distortions of the image \cite{demene2021transcranial,robin2023vivo, xing2023phase}. To alleviate the impact of aberrations, a craniotomy is often performed in pre-clinical studies\cite{mace2011functional, mace2013functional, errico2015ultrafast} but limit their scope and the clinical potential of ultrasound-based brain imaging techniques.

Numerous methods have been proposed in the past decades for aberration correction\cite{o1988phase, flax1988phase, maasoy2005iteration,montaldo2011time, osmanski2012aberration}, some specifically for transcranial imaging applications \cite{ivancevich2008real,lindsey20143, soulioti2019super, schoen2019heterogeneous, tian2023transcranial}. Since the distortion of the wave fronts of the reflected echoes are directly related to the aberration function, the use of single point-like reflectors or diffuse scatterers has been proposed to perform phase aberration correction \cite{o1988phase, flax1988phase}.

 \begin{figure}[t]
    \centering
    \includegraphics[width=1\linewidth]{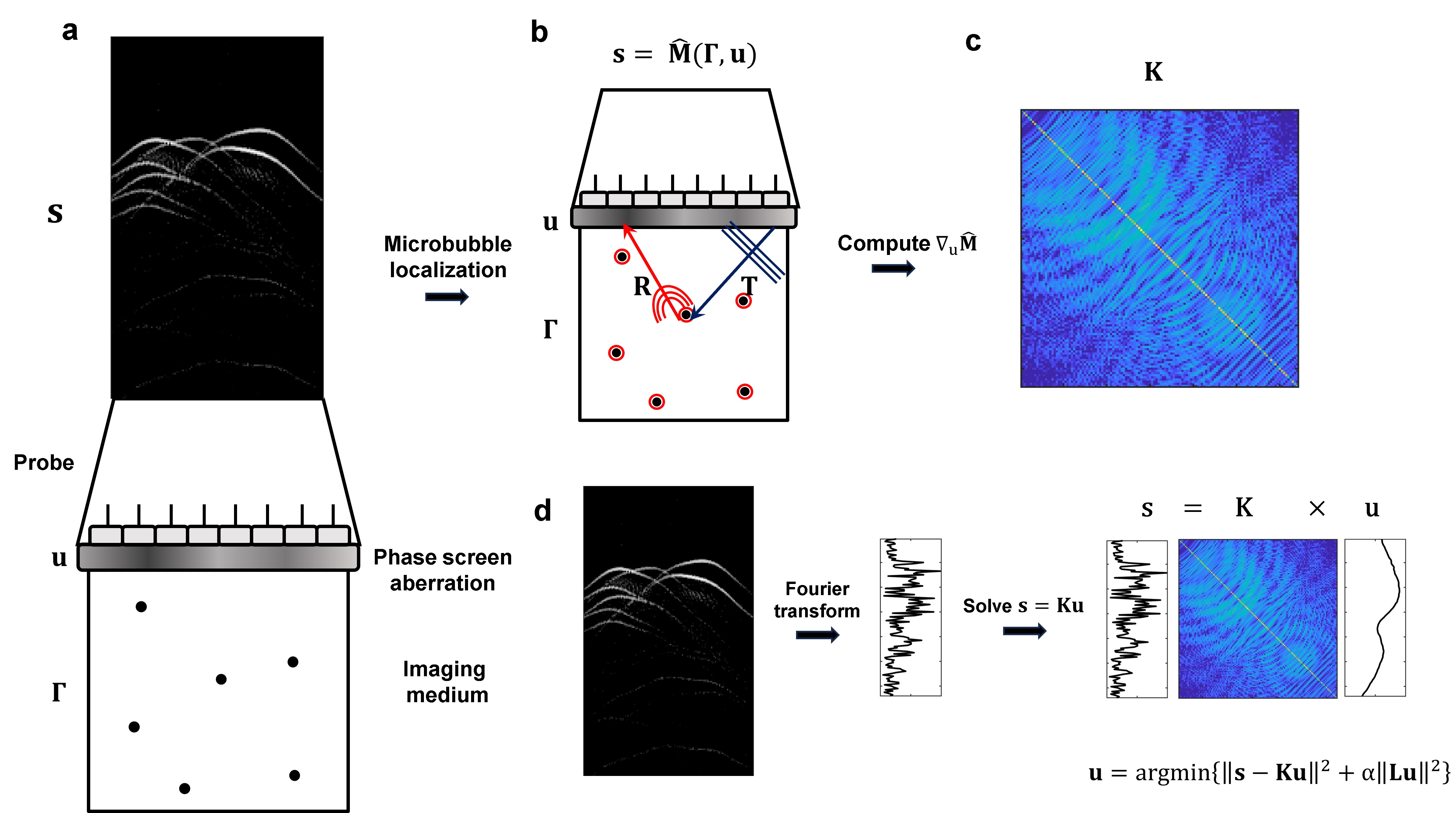}
    \caption{Schematic representation of the inverse problem approach. (a) A linear probe is used to image a scattering medium  $\mathbf{\Gamma}$ in presence of a phase screen aberrator $\mathbf{u}$. The acquired signal is then given by $\mathbf{s}$. (b)  $\mathbf{\Gamma}$ can be inferred from microbubble positions. A forward model $\hat{\mathbf{M}}$ based on wave propagation can then be built, linking the medium $\mathbf{\Gamma}$ and aberration $\mathbf{u}$ to the measured signal  $\mathbf{s}$. The transmit operator $\mathbf{T}$ is used to link the transmitted field of the probe to each scatterer. The  reflection operator  $\mathbf{R}$ then links the scatterer signals back to each element of the probe. (c) Computing the gradient term  $\nabla_{\mathbf{u}}\hat{\mathbf{M}}$ allows to write the forward model in matrix form $\mathbf{K}$ (d). In the Fourier domain, the forward model with the matrix $\mathbf{K}$ links directly the received signal $\mathbf{s}$ to the aberration function $\mathbf{u}$ in a linear form. Applying inversion strategies such as least-square minimization is used to solve $\mathbf{u}$.}
    \label{fig:model}
\end{figure}

Highly echoic microbubbles were first invented for contrast enhance ultrasound (CEUS) imaging and were later used for other ultrafast ultrasound techniques including ULM \cite{christensen2014vivo, errico2015ultrafast}. These microbubbles have been used for imaging through the skull since their signals compensate for the attenuation, as well as targets for performing correction by isolating single microbubble hyperbolas to retrieve the aberration function \cite{oreilly2013super, soulioti2019super, demene2021transcranial, robin2023vivo}. However, the application of previous strategies is complicated by the presence of multiple microbubbles with crossing signals, which cause interference that disrupt hyperbolas and coherence profiles when microbubbles are not adequately isolated\cite{oreilly2013super}. To alleviate those issues, virtual focusing and directional filtering were proposed to reject signal originating from other microbubbles\cite{robin2023vivo}. \textit{In vivo} microbubble signals are also limited by the directivity of the probe elements, especially at higher frequency, meaning that the diffraction hyperbola is not entirely defined along the aperture. Consequently, aberration correction methods are not commonly applied in CEUS or ULM. Deep learning strategies have been proposed to solve some of these issues, which outperformed conventional methods at higher microbubble concentrations \cite{xing2023phase}.

 
Herein, we propose a model-based optimization approach for phase aberration correction suitable for ultrafast imaging that leverages more specifically the sparsity of the entire backscattered signals of microbubbles in CEUS and ULM without requiring any change in the transmit scheme. Contrary to previous microbubble-based correction method, this inverse problem approach to aberration correction (IPAC) takes into account a larger amount of information in the ultrasound signals, such as probe and pulse characteristics, as well as the physics of wave interaction and interference coming from multiple scatterers, averting the need for physically or virtually isolated "guided stars". IPAC can be generalized to any ultrasound probe and is compatible for plane waves and divergent waves transmission. We first describe the physical and mathematical formalism of the forward model. We then describe how this model can be expressed into a linear form by defining the analytical expression the gradient operator with respect to the aberration function. We used simulation studies to validate the efficiency of IPAC to retrieve the aberration function. Finally, we validated on \textit{in vivo} transcranial data acquired on five mouse brains on both Power Doppler and ULM images.

\begin{figure*}[ht!]
    \centering
    \includegraphics[width=1\linewidth]{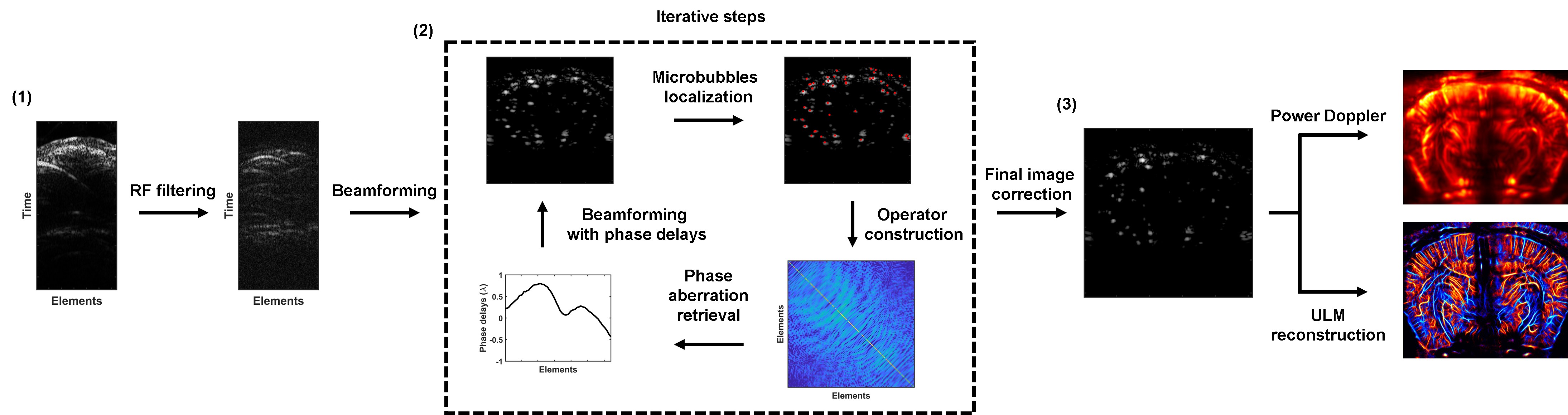}
    \caption{General framework for \textit{in vivo} phase aberration correction. (1) A SVD filter is applied to raw RF channel data to isolate microbubble hyperbolas. (2) Microbubble positions are localized on beamformed data to construct the forward model $\mathbf{K}$. The aberration function is then retrieved by using a least-squares minimization. Delays associated to the aberration function are included into the beamformer to perform correction. This step can be performed in an iterative manner to correct for microbubble positions, which return a more precise  $\mathbf{K}$ matrix. (3) After convergence of the iterative step, the final aberration function is used to perform correction on either power Doppler or ULM images.}
    \label{fig:framework}
\end{figure*}
\section{Theory and mathematical description}

\subsection{Theoretical description of the forward problem}

For low concentrations, CEUS images can be represented as a sparse distribution of scatterers. A sparse representation is also present in the ultrasound radiofrequency (RF) channel signals as spatio-temporally separated echoes, i.e., diffraction hyperbolas. Since microbubbles act as point-like scatterers, they yield the Green function of the imaging system associated to their given position. A forward model $\hat{\mathbf{M}}$ can be developed to link the sparse measured signals $\mathbf{s}$ (i.e., the received acoustic RF channel data) to the sparse medium  $\mathbf{\Gamma}$ (i.e., the image)
\begin{align}
    \mathbf{s}= \hat{\mathbf{M}}(\mathbf{\Gamma}).
\end{align}
Since $\hat{\mathbf{M}}$ is here expressed as generally as possible, it can be used to model any kind of ultrasound probe. 

In the Fourier domain, assuming linear propagation, this forward model can be defined as the product between matrix operators describing the physics of wave transmission, scattering, and reflection. To demonstrate this, let us consider the case of a one-dimensional ultrasound probe composed of $N_{e}$ elements where each element is located at the position $\mathbf{r}_n=(x_n,y_n,z_n)$, with $x_n$, $y_n$, and $z_n$ representing the coordinates of the element. This probe is used to insonify a medium composed of $N_s$ scatterers that is defined by the operator $\mathbf{\Gamma}$. Upon linear and single scattering assumption, $\mathbf{\Gamma}$ can be described in the form of a diagonal matrix\cite{lambert2020reflection} and its coefficients $\Gamma_{s,s}(\omega) = \Gamma(\mathbf{r}_s,\omega)$ represent the backscattering amplitude of each scatterer at position $\mathbf{r}_s=(x_s,y_s,z_s)$ for the wave angular frequency $\omega=2\pi f$, with $x_s$, $y_s$, and $z_s$ the coordinates of the scatterer. The transmission operator $\mathbf{T}$ with coefficients $T_{n,s}(\omega)=T(\mathbf{r}_n,\mathbf{r}_s,\omega)$ then links the transmitted field of each element at the position $\mathbf{r}_n$ to any scatterer at the position $\mathbf{r}_s$ within the medium. Likewise, the coefficients $R_{m,s}(\omega)=R(\mathbf{r}_m,\mathbf{r}_s,\omega)$ of the reflection operator $\mathbf{R}$ describe the reflection from any point at the position $\mathbf{r}_s$ back to each receiving element at the position $\mathbf{r}_m$ (see figure \ref{fig:model}).

Considering the transmission of all $N_e$ emitting elements and the contribution of all $N_s$ scatterers within the medium, the model signal $\hat{M}_m$ received by the $m^{th}$ element of the probe can be defined as described by \cite{garcia2022simus}
\begin{align}
\hat{M}_{m}(\mathbf{\Gamma}, \omega)=\sum_{s=1}^{N_s}\sum_{n=1}^{N_{e}}R_{m,s}(\omega)\Gamma_{s,s}(\omega)T_{n,s}(\omega).
\end{align}

For a detailed description of the theoretical derivation of the  transmission and reflection operators, see the \nameref{section:appendix}. Furthermore, the skull-induced aberration $\mathbf{u}$ can also be included into this forward model
\begin{align}
    \mathbf{s}= \hat{\mathbf{M}}(\mathbf{\Gamma}, \mathbf{u}).
\end{align}

Again, $\mathbf{u}$ is here expressed as generally as possible to model any kind of aberration. Let us now consider the specific case of a near-field phase screen aberrator placed between the medium and the probe. This aberration model can be described as time delays associated to the transmitted and received fields for each element of the ultrasound probe while the medium is considered globally uniform, as previously described \cite{flax1988phase, o1988phase}. In mathematical form, the aberration profile can be defined as a function of each element
\begin{align}
u_{n}(\omega)=u(\mathbf{r}_n,\omega)=a({\mathbf{r}_n})e^{i\omega \tau(\mathbf{r}_n)},
\end{align}

where $\tau(\mathbf{r}_n)$ stands for the phase delay and $a(\mathbf{r}_n)$ the amplitude aberration associated to the $n^{th} $ element. The complete aberration function can hence be defined as the vector $\mathbf{u}(\omega)$. The RF channel signals in presence of phase aberration in both transmit and receive can be modeled as
\begin{align}
\label{eq:pressure_ab}
\hat{M}_{m}(\mathbf{\Gamma}, \mathbf{u_{}},\omega)&=\sum_{s=1}^{N_s}\sum_{n=1}^{N_{e}}u_{m}(\omega)R_{m,s}(\omega)\Gamma_{s,s}(\omega)T_{n,s}(\omega)u_{n}(\omega) .
\end{align}

Here, we assumed that the same aberration function is present both in transmit and receive.

\subsection{Linear formulation of the forward problem}

We have considered previously in our model the signal received by a single element. For sake of clarity, if we consider only the central frequency $\omega=\omega_0$ and by using the operators $\mathbf{T}$, $\mathbf{\Gamma}$ and $\mathbf{R}$, we can write equation (\ref{eq:pressure_ab}) in a matrix form that fully describes the forward model
\begin{align}
    \mathbf{\hat{M}(\mathbf{\Gamma},u)} =\mathbf{u}\circ\mathbf{H}\mathbf{u},
    \label{eq:compact_vec_form}
\end{align}

where the symbol $\circ$ stands for the Hadamard product and
\begin{align}
     \mathbf{H}= \mathbf{R}\times\mathbf{\Gamma}\times\mathbf{T}^T.
     \label{eq:gradient}
\end{align}

Here $\mathbf{H}$ is the transfer function of the imaging system that describes all the physics of the wave propagation from each element in transmit to each element in receive. Hence,  $\mathbf{H}$ can also be seen as the wave propagator operator similarly to what was previously described \cite{tanter2000time, tanter2001optimal}. The particularity of our formalism is that $\mathbf{H}$ is written in such a way that it is uncoupled from the aberration function, instead of the probe emission. Moreover, considering the identity $\mathbf{x}\circ \mathbf{y} = diag(\mathbf{x})\mathbf{y}=diag(\mathbf{y})\mathbf{x}$, the differential of $\hat{\mathbf{M}}$ is given by
\begin{align}
    d\hat{\mathbf{M}} = diag(\mathbf{H}\mathbf{u}) d\mathbf{u} +  diag(\mathbf{u}) d(\mathbf{H}\mathbf{u}).
\end{align}

We can then express the gradient of $\hat{\mathbf{M}}$ with respect to $\mathbf{u}$ as
\begin{align}
\nabla_{\mathbf{u}}\hat{\mathbf{M}} (\mathbf{\Gamma}, \mathbf{u})    &= diag(\mathbf{H}\mathbf{u})\nabla_{\mathbf{u}}\mathbf{u} + diag(\mathbf{u})\nabla_{\mathbf{u}}(\mathbf{H}\mathbf{u})    
\end{align}

which simplifies as 
\begin{align}
\nabla_{\mathbf{u}}\hat{\mathbf{M}} (\mathbf{\Gamma}, \mathbf{u})    = diag(\mathbf{H}\mathbf{u})+diag(\mathbf{u})\mathbf{H}.
\label{eq:frechet}
\end{align}

 The gradient operator $\nabla_{\mathbf{u}}\hat{\mathbf{M}}$ is  a $ N_{e}\times N_{e}$ tensor that describes the variation of the received signal for each element of the ultrasound probe caused by the aberration. By introducing  $\nabla_{\mathbf{u}}\hat{\mathbf{M}}$, the forward model can be approximated in a linear form by using a Taylor expansion of the first order
 \begin{align}
\mathbf{s}&\approx \hat{\mathbf{M}}(\mathbf{\Gamma}, \mathbf{u}_{0})+\nabla_{\mathbf{u}}\hat{\mathbf{M}}(\mathbf{\Gamma}, \mathbf{u}_{0})(\mathbf{u}-\mathbf{u}_{0} ),
\label{eq:taylor}
\end{align}
with $\mathbf{u}_{0}$  an arbitrary aberration function, which can be used as an initial guess later on. In practice, $\mathbf{u}_0$ can be first set as the unitary vector $\mathbf{1}^{N_{e}\times 1}$, and then $\hat{\mathbf{M}}(\mathbf{\Gamma}, \mathbf{u}_0)$ represents the theoretical expected signal in absence of aberration. Once the problem is written into this formalism, it becomes linear with respect of $\mathbf{u}$, meaning that a direct forward operator  $\mathbf{K}$ can be represented as a  matrix which links the measured signals to the aberration function, such as
\begin{align}
\mathbf{s}\approx\mathbf{K}_{}(\mathbf{\Gamma}, \mathbf{u}_0){\mathbf{u}},
\end{align}
with
\begin{subequations}
\begin{align}
\mathbf{K}(\mathbf{\Gamma}, \mathbf{u}_0)&=\Big(\nabla_{\mathbf{u}}\hat{\mathbf{M}}(\mathbf{\Gamma}, \mathbf{u}_0)\Big| \hat{\mathbf{M}}(\mathbf{\Gamma}, \mathbf{u}_0) - \nabla_{\mathbf{u}}\hat{\mathbf{M}}(\mathbf{\Gamma}, \mathbf{u}_0)\mathbf{u}_0 \Big)\\
{\mathbf{u}}&:=\begin{pmatrix}
\mathbf{u}\Big|1
\end{pmatrix}^T.
\end{align}
\end{subequations}

\textit{A priori} knowledge of the medium based on the detection of microbubble positions can be used to build the forward model matrix $\mathbf{K}$.  Application of inversion strategies that minimize the mismatch between the measurements $ \mathbf{s}$ and the model  $\mathbf{K}$ can be used to retrieve $\mathbf{u}$ and then perform image correction by including the appropriate delays into the beamforming process.

\subsection{Least-squares solution to the inverse problem}

Inverse problems are usually ill-posed and sensitive to noise. A robust estimation of the aberration function can be retrieved by using a least-squares method and a Tikhonov regularization\cite{tihonov1963solution} 
\begin{align}
\hat{\mathbf{u}}=\text{argmin}\{ ||\mathbf{s}- \mathbf{K}(\mathbf{\Gamma}, \mathbf{u}_0){\mathbf{u}}||^2 + \alpha ||\mathbf{L}{\mathbf{u}} || ^2\},
\end{align}

with $\hat{\mathbf{u}}$ the least-squares solution,  $\mathbf{L}$  the regularization matrix and $\alpha$ a regularization parameter. For efficient noise reduction and to retrieve smooth and continuous solutions for the aberration function, the $\mathbf{L}$ matrix can be set as the second $\mathbf{D}_2$ derivative matrix. In matrix form, $\hat{\mathbf{u}}$ can be written as
\begin{align}
\hat{\mathbf{u}}=(\mathbf{K}^\dag(\mathbf{\Gamma}, \mathbf{u}_0)\mathbf{K}(\mathbf{\Gamma}, \mathbf{u}_0)+\alpha \mathbf{D_2}^T\mathbf{D_2})^{-1}\mathbf{K}^\dag(\mathbf{\Gamma}, \mathbf{u}_0)\mathbf{s}.
\label{eq:ls_solution}
\end{align}

The solution $\hat{\mathbf{u}}$ represents a first approximation of the aberration function, which can be used to adjust delays in beamforming to better reconstruct the image and improve microbubble localization to correct $\mathbf{\Gamma}$. Our formalism can then be used in an iterative manner in order to achieve a more precise computation of the aberration function. Indeed, by updating the initial guess state $\mathbf{u}_0$
\begin{align*}
\mathbf{u}_0:= \hat{\mathbf{u}},
\end{align*}
one can further update the operator $\mathbf{K}$ (by using the updated values of  $\mathbf{\Gamma}$ and $\mathbf{u}_0$) which would in turn gives a more precise estimation of $\mathbf{u_{}}$. This iterative process can be applied for $n$ iterations or until reaching a convergence criterion, such as $\lVert \hat{\mathbf{u}}- \mathbf{u}_0\rVert < \epsilon$, $\epsilon$ being the tolerance value.

\subsection{Generalization to multiple transmit angles}

Until now, we have considered only a single transmit angle. In ultrafast imaging, multiple transmit angles $\theta$ are used to improve resolution and increase signal contrast\cite{montaldo2009coherent} and signal-to-noise ratio (SNR) \cite{tiran2015multiplane}. The receive signal $\mathbf{s_\theta}$ would then be a matrix of dimension $N_{e}\times N_\theta$, with $N_\theta$ the number of transmit angles. Within our formalism, this requires to consider an additional phase delay in transmission for each element $A_n(\omega,\theta)=e^{i\omega\tau_n(\theta)}$. We can define a transmit operator $ \mathbf{T}_\theta$ as a tensor of dimensions $N_{e}\times N_s \times N_\theta$ that includes each transmit angle
\begin{align}
    \mathbf{T}_\theta = \mathbf{T}\circ \mathbf{A}(\theta).
\end{align}

Assuming the aberration function defined in the transducer basis does not change with the transmit angle, the problem can be reduced back into a two-dimensional system. Indeed, by vectorizing $\mathbf{s}_\theta$ and reshaping  $\mathbf{M}_\theta$, we can write
\begin{align}
\mathbf{s}_\theta= \mathbf{K}_\theta(\mathbf{\Gamma}, \mathbf{u}_0){\mathbf{u}},
\end{align}

with
\begin{align}
\mathbf{K}_\theta(\mathbf{\Gamma}, \mathbf{u}_0)
&=
\begin{bmatrix}
\mathbf{K}_{1}(\mathbf{\Gamma}, \mathbf{u}_0)\\
\vdots\\
\mathbf{K}_{N_\theta}(\mathbf{\Gamma}, \mathbf{u}_0)
\end{bmatrix},
\end{align}

and thus the solution takes the same form as equation \ref{eq:ls_solution}.

\begin{figure*}[t]
    \centering
\includegraphics[width=0.7\linewidth]{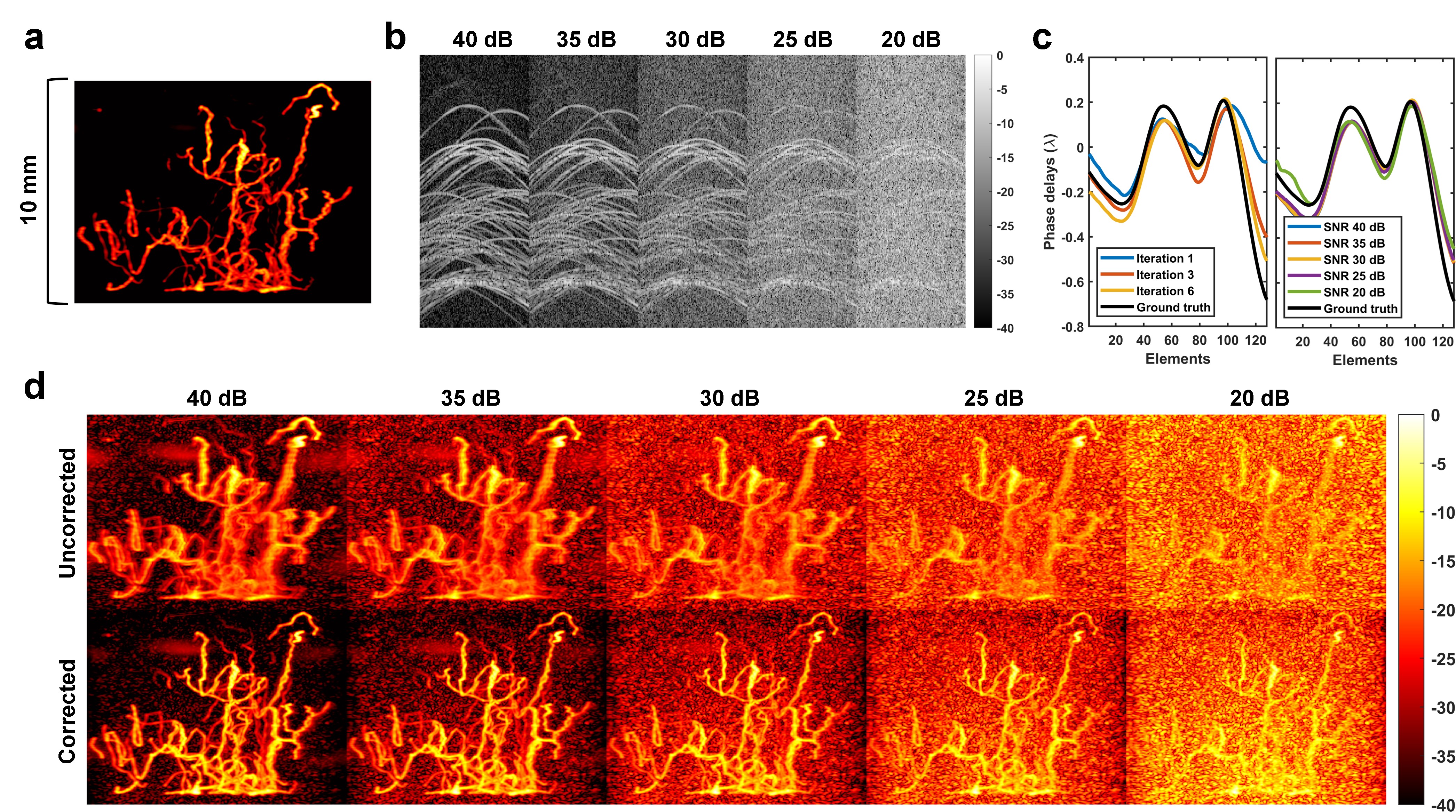}
\caption{Aberration correction on \textit{in silico} data using a linear probe. (a) Simulation ground truth of the vascular network. (b) Simulated magnitude of RF channel signals with aberration for different SNR. (c) Aberration functions retrieved with the correction method for different SNR and iterative steps. (d) Comparison of Power Doppler images before and after correction for different SNR.}
    \label{fig:in_silico}
\end{figure*}
\begin{figure*}[t]
    \centering
\includegraphics[width=0.7\linewidth]{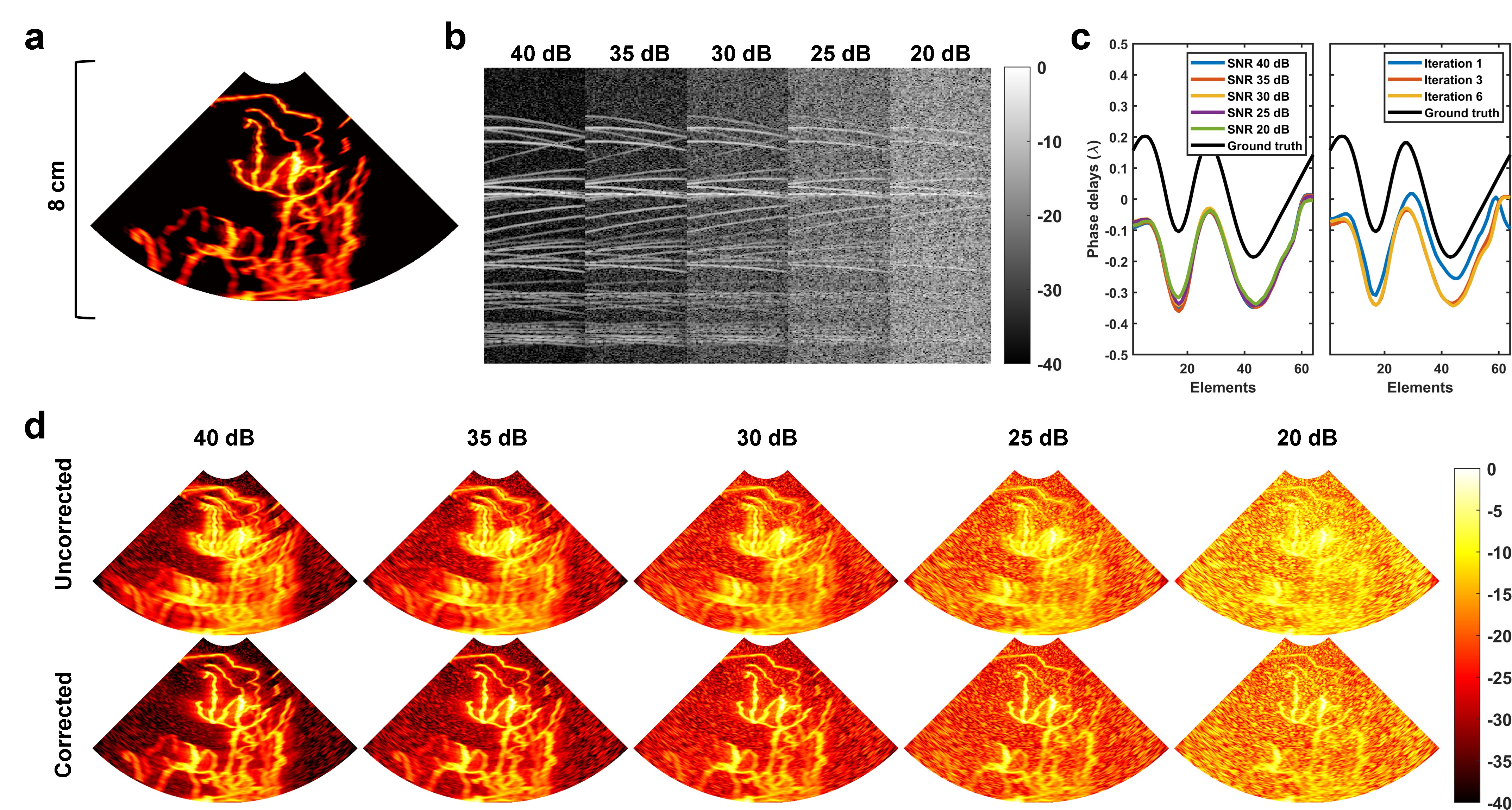}
\caption{Aberration correction on \textit{in silico} data using a phased array probe. (a) Simulation ground truth of the vascular network. (b) Simulated magnitude of RF channel signals with aberration for different SNR. (c) Aberration functions retrieved with the correction method for different SNR and iterative steps. (d) Comparison of Power Doppler images before and after correction for different SNR.}
    \label{fig:in_silico_phased_array}
\end{figure*}

\section{Methods}

\subsection{\textit{In silico} experiment}

The \textit{in silico} datasets were simulated using a GPU implementation of the SIMUS software \cite{garcia2022simus}, which simulates the pressure fields in the frequency domain for each probe element in transmission and reception. SIMUS was modified to include the phase screen aberration model for both the emission and reception field simulations. Each plane wave was simulated by time-delaying each element during transmission by the corresponding angle. A uniform distribution was used to generate random aberration profiles, both in amplitude and phase. The amplitude attenuation (which models the skull attenuation) ranged from 0-50$\%$ and the relative phase variation was limited to 1 wavelength ($\lambda$).

To simulate the motion of microbubbles, we used a previously described \cite{milecki2021deep, belgharbi2023anatomically}, graph-based model of the vasculature of a mouse brain with Poiseuille flow conditions. The vascular network was dilated to cover a field of view comparable to \textit{in vivo} studies. The number of microbubbles per frame was set between 50 and 100 to reproduce  concentrations comparable to the ones used in ULM experiments \cite{heiles2022performance}. Simulated microbubbles were treated as point-like scatterers. The ultrasound radiofrequency (RF) response for each microbubble position was simulated for 11 compounding angles (-5$^\circ$ to 5$^\circ$ with a 1$^\circ$ increment) and a 128-element, 15.625-MHz linear array transducer ultrasound probe with the properties of the L22-14 (Vermon, France) used during \textit{in vivo} experiments. To demonstrate the capability of our correction method to generalize to any kind of 1D ultrasound probe, we also simulated a 64-element, 2.5-MHz phased array with the properties of the P4-2 (Phillips, USA).  The speed of sound used for all simulation was set to 1540 m/s.
 
Fully developed speckle and electronic noise were added to every simulation to emulate realistic ultrasound acquisition. The amplitude of the noise was set to different level of SNR between 40 dB and 20 dB. To emulate the Verasonics Vantage system processing, the RF channel data were subsampled into 100$\%$ bandwidth in phase-quadrature (IQ) data.

\subsection{\textit{In vivo} experiment}

Ultrasound acquisitions were performed with the Vantage system (Verasonics Inc., Redmond, WA) with a 100$\%$ sampling rate.  A 15.625-MHz probe (L22-14, Vermon, France) and 11 compounding angles (-5$^\circ$ to 5$^\circ$ with step of 1$^\circ$) were used for the acquisitions made on five 8-12 week-old mice. Excess hair was removed from the head, with the skin and skull kept intact through all the experiments. A 3D-printed head mount was used to limit movements from the mouse head and ultrasound reflection. The mice were kept under general anesthesia with Ketamine (50 mg/kg) and Medetomidine (1 mg/kg) and body temperature maintained between 36 and 37 °C during the acquisition. Definity microbubbles (Perflutren Lipid Microsphere, Lantheus Medical Imaging, Billerica, MA, USA) injection was performed in the retro-orbital sinus with a 10$\mu$L bolus. The mice were imaged at different coronal planes identified as Bregma -2.2 mm, -1.5 mm, 0.0 mm, +0.2 mm, and +0.7 mm. A complete acquisition consisted of 600 buffers composed of 500 images acquired at 1000 frames per second (mice imaged at Bregma -2.2 mm, -1.5 mm, and 0.0 mm) or 1200 buffers of 400 images acquired at 1600 frames per second (mice imaged at Bregma +0.2 mm and +0.7 mm), for a total acquisition time of 5 minutes. The pulse shape was composed of three cycles of a sinusoidal wave emitted at 25 V. The mechanical index (MI) at elevation focus was 0.20.
All experimental procedures were approved by the Animal Care Ethics Committee of the University of Montreal (Permit Number: 21-017 and 22-013).

A singular value decomposition (SVD) clutter filter\cite{demene2015spatiotemporal} was used on raw ultrasound data prior to beamforming to isolate microbubble hyperbolas\cite{robin2023vivo, yan2023fast} by reshaping the signal into a spatio-temporal Casorati  matrix of dimensions $N_sN_{e}N_\theta\times N_f$, with $N_f$ the number of frames. The SVD threshold was set heuristically to maximize tissue rejection. An additional bandpass filter was applied with cutoff frequencies empirically set to optimize noise reduction. Raw ultrasound data were beamformed a first time with the DAS beamformer and microbubble positions were retrieved using a previously published localization algorithm \cite{milecki2021deep, bourquin2021vivo}.

\begin{figure}
    \centering
\includegraphics[width=1.0\linewidth]{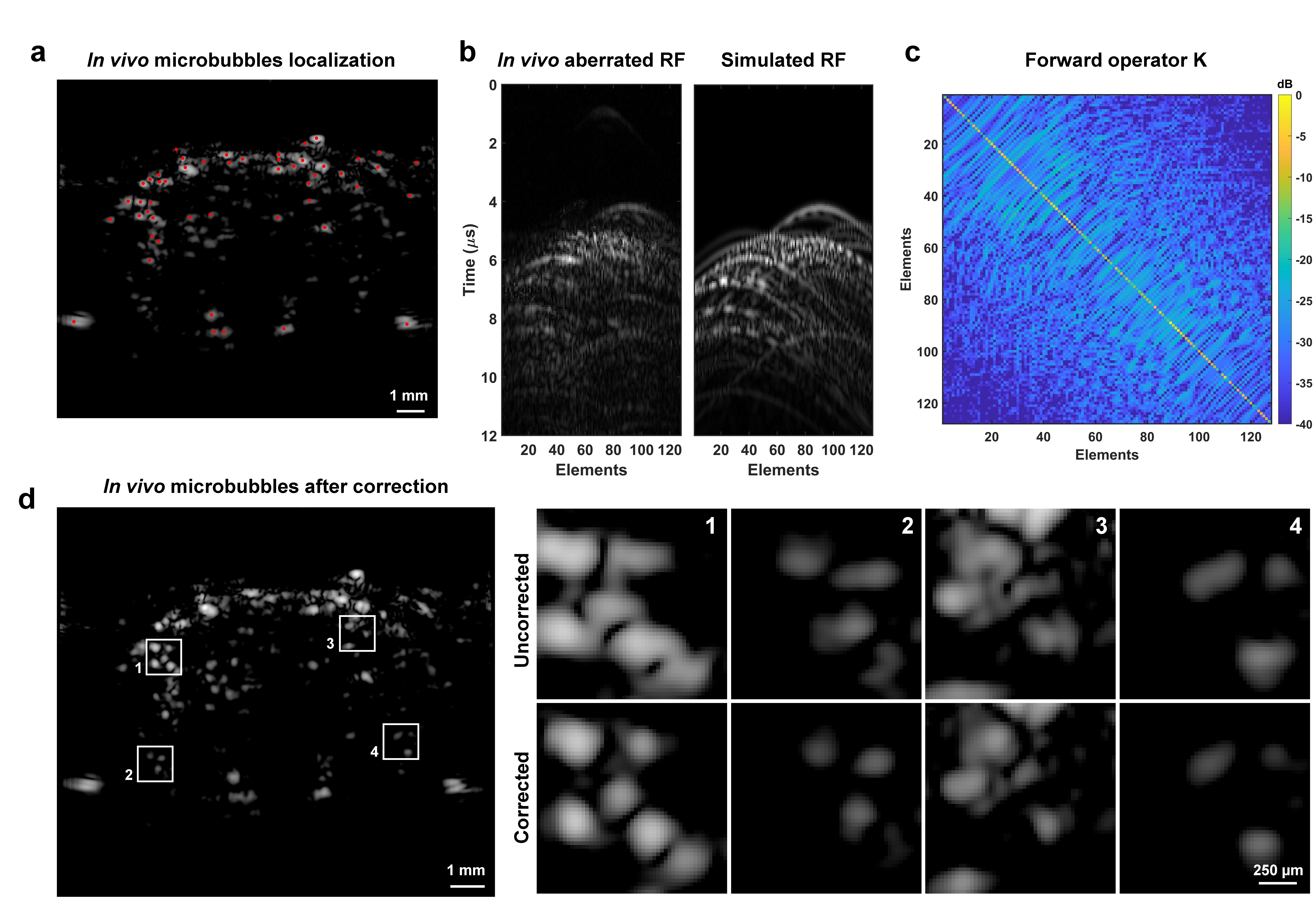}
    \caption{Example of \textit{in vivo} microbubble signals used for aberration correction. (a) Microbubbles detection and localization on a mouse brain. (b) \text{In vivo} RF channel signals compared to the simulated magnitude of RF channel signals with the forward model. (c) Forward model matrix $\mathbf{K}$ associated to the RF channel data. (d) Microbubbles after correction.}
    \label{fig:invivo_microbubble}
\end{figure}

\begin{figure*}[ht]
    \centering
    \includegraphics[width=1\linewidth]{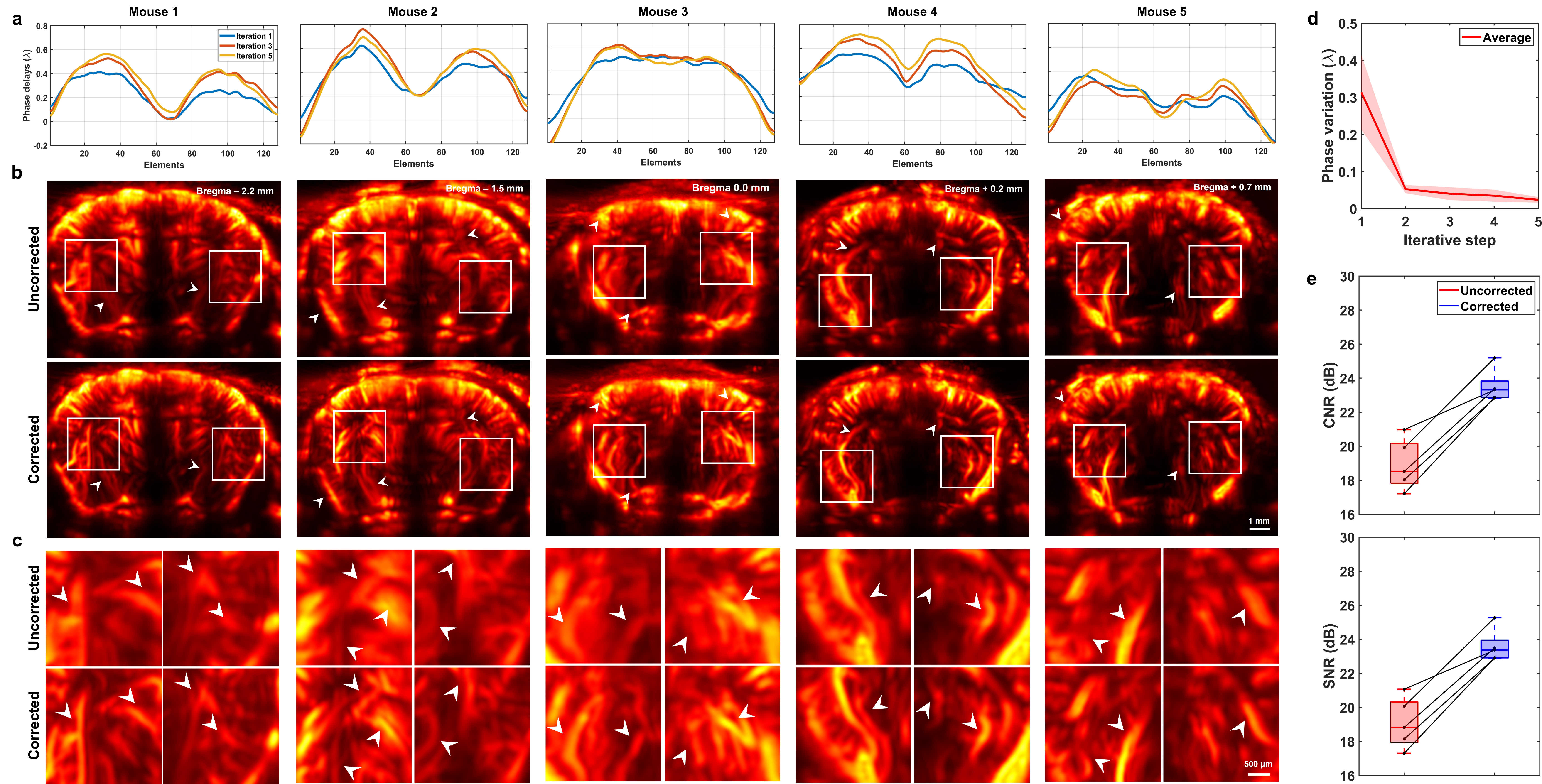}
    \caption{\textit{In vivo} aberration correction on transcranial CEUS Power Doppler images of 5 different mouse brains. (a) Aberration function profile of each mouse. After 5 iterations, most aberration profiles reached stability. (b) Comparison of the brain vasculature before and after aberration correction. (c) Close-up regions extracted from (b). (d) Mean phase variation of the aberration function between each iterative step. (e) CNR is improved after correction (one-tailed paired t-test, p $<$ 0.0001). SNR is improved after correction (one-tailed paired t-test, p $<$ 0.0001).}
    \label{fig:doppler}
\end{figure*}

\subsection{Construction of the model matrix \texorpdfstring{$\mathbf{K}_\theta$}{Lg}}

 In order to consider only intense microbubbles and reduce false detections, a 20 dB threshold was used on detected microbubbles. The model matrix $\mathbf{K}_\theta$ was built directly in the Fourier domain using information on the bandwidth and element geometry of the different ultrasound probes provided by the Verasonics system. The same pulse definition used during \textit{in vivo} experiment was used to define $\mathbf{T}_\theta$. Further details on the equations are given in the \nameref{section:appendix} and in reference \cite{garcia2022simus}. $\mathbf{\Gamma}$ was inferred from detected microbubble positions and backscattering amplitudes using a 2D Gaussian fitting as described in section \ref{section:ULM}. The regularization term was set heuristically to 0.01 which retrieved continuous and smooth aberration profiles.

\subsection{Aberration correction method}
The retrieved aberration function in the Fourier domain consisted of a matrix of dimension $N_e\times N_\omega$. Both amplitude and phase aberration were computed as part of the least-squares solution, but only the phase was kept in the subsequent correction process. A 2D phase unwrapping method robust to noise \cite{zhao2018robust} was applied to retrieve the time delays. An averaging of the time delays retrieved over different acquisition frames was made. Aberration correction was performed by including the time delays into the beamformer for each angle before compounding. The same aberration function was applied for all frames. To reduce errors caused by precision on microbubble localization, we performed correction in an iterative manner. The images were beamformed with the time delays and microbubble detected again to build a more precise $\mathbf{K}_\theta$, with the previously mentioned step applied again. The entire correction pipeline is illustrated in figure \ref{fig:framework}.

\subsection{Coherence-based method}
The coherence-based aberration correction method \cite{o1988phase, demene2021transcranial, robin2023vivo} was adapted for comparison purpose. Channel data were filtered as previously described to reveal microbubble signals. hyperbolas of individual microbubbles were rephased by shifting each channel by the time delays $t_n$ computed accordingly to their positions $(x_s,z_s)$

\begin{align}
    t_n(x_s,z_s)= \frac{z_s\cos\theta+x_s\sin\theta+\sqrt{z_s^2+(x_s-x_n)^2}}{c},
\end{align}

where $x_n$ represents the position of each element, $\theta$ the transmit angle, and $c$ the speed of sound. The phase delays associated to the aberration profile were calculated by finding the maximum of the cross-correlation between each element in the Fourier domain with zero paddings.

\subsection{ULM reconstruction}\label{section:ULM}

Positions of individual microbubbles were detected by using a 2D Gaussian fitting. Microbubbles with correlation with the point spread function (PSF) higher than 0.5 were then tracked using an algorithm based on the Hungarian method (\url{https://github.com/tinevez/simpletracker}) with no gap filling and maximal linking distance of 2 pixels. A smoothing algorithm based on least-squares methods \cite{garcia2010robust} was applied on each track. Tracks were then interpolated using the modified Akima method \cite{akima1970new} to retrieve continuous trajectories and projected  using the maximal intensity projection on a factor 10 interpolation grid \cite{heiles2022performance}. Velocities were computed with a forward finite difference scheme directly after smoothing and then interpolated in a similar manner. Only tracks longer than 15 frames were used for the ULM reconstruction. A 2D Gaussian filtering with standard deviation of 1 pixel was used to reduce noise and improve ULM map rendering.

\subsection{Image quality assessment}

The improvement of Power Doppler images after correction was determined by evaluating the contrast-to-noise ratio (CNR) and SNR
\begin{align}
    CNR &= 10 \log_{10}\left(\frac{|\mu_{vessel}-\mu_{noise}|}{\sigma_{noise}}\right)\\
    SNR &= 10\log_{10} \left(\frac{\mu_{vessel}}{\sigma_{noise}}\right),
\end{align}

where $\mu_{vessel}$ is the mean intensity in a region of interest within a large vessel,  $\mu_{noise}$, and $\sigma_{noise}$ the mean intensity and standard deviation within a region of interest  defined in the background, respectively.

Similarity between aberration functions $\mathbf{u}_1$ and $\mathbf{u}_2$ retrieved at different concentration was evaluated with a modified cosine similarity
\begin{align}
    C = \frac{\mathbf{u}_1 \cdot \mathbf{u}_2}{\max{\left(||\mathbf{u}_1||^2, ||\mathbf{u}_2||^2\right)}},
\end{align}

where $\cdot$ represents the dot product. Here, the maximal norm of $\mathbf{u}_1$ or $\mathbf{u}_2$ was used as normalization to account for scale difference between $\mathbf{u}_1$and $\mathbf{u}_2$.

For ULM images, the Fourier ring correlation (FRC) \cite{hingot2021measuring} was used to evaluate the gain in resolution after correction by adapting code from \url{https://github.com/bionanoimaging/cellSTORM-MATLAB}. Tracks were randomly separated into two sub-images. The FRC was computed by using correlation between the spatial spectrum of each sub-image for pixels within a radius $r$
\begin{align}
    FRC(r) = \frac{\sum_{r\in R}F_1(r)\cdot F_2(r)^*}{\sqrt{\sum_{r\in R}|F_1(r)|^2\cdot \sum_{r\in R}|F_2(r)|^2}}.
\end{align}

The intersection of the FRC curve with the half-bit threshold was used to establish the resolution.

\subsection{Statistical analysis and reproducibility}

Reproducibility of IPAC was evaluated in 5 different mice. Statistical significance of SNR and CNR improvement of CEUS images were evaluated with a parametric one-tailed paired t-test with a 95$\%$ confidence level assuming a normal distribution of the data and using the before and after correction values as paired measurements. Likewise,  statistical significance of mean microbubble correlation, mean track length, number of tracks, mean velocity, and resolution of ULM images were evaluated with a  parametric one-tailed paired t-test. Multiple groups comparison for microbubble concentration analysis were first evaluated with a one-way ANOVA (analysis of variance). If applicable, a post hoc pairwise comparison using a paired t-test was used to identify which groups were statistically different.

The nonparametric Kolmogorov–Smirnov (KS) test was used to assess statistical difference between distributions of microbubble correlations, track lengths, and track velocities, without any assumption on the data distribution. The Watson-Williams test for circular data was used to assert significance on flow direction. A p-value of $<0.05$ was considered statically significant. Levels of significance are given as * : p$< 0.05$, ** : p$< 0.01$, *** : p$< 0.001$, and **** : p$< 0.0001$.

\subsection{Computational resources}

Processing for aberration correction was performed in MATLAB R2023b on a working station with an Intel Xeon W-2245 3.9 GHz processor, 256 Go of RAM, and an NVIDIA GeForce RTX 3090 GPU with 24 Go of dedicated memory. Codes for generating the matrix $\mathbf{K}$ and computing the aberration function were implemented in the MATLAB language. Generating the matrix $\mathbf{K}$ and computing the inversion algorithm for a single frame took an average time of 13.5 seconds. The beamformer was fully implemented on GPU using CUDA. Including the time delays into the beamformer did not significantly change beamforming processing times, which were on average 6.6 seconds for a single buffer. Processing a single CEUS Power Doppler image (beamforming and SVD filtering) took an average time of 8.5 seconds.

ULM processing (beamforming, microbubble localization and tracking) was performed on high-performance compute (HPC) system. Beamforming and localization were performed on GPU system using an Intel Gold 6148 Skylake 2.4 GHz processor, 64 Go of RAM, and an Nvidia V100SXM2 GPU with 16 Go of dedicated memory.  Buffers were processed in parallel in less than 1 hour using multiple computing nodes. Loading and processing a single buffer took an average time of 45 seconds. Only detected microbubble positions and correlation values were saved. Tracking was performed separately on CPU only system using 10 cores (Intel Gold 6148 Skylake 2.4 GHz processor) and 128 Gb of RAM. Buffers were processed in parallel on workers and generating a complete ULM map took less than 20 minutes.

\begin{figure}[t!]
    \centering
    \includegraphics[width=1\linewidth]{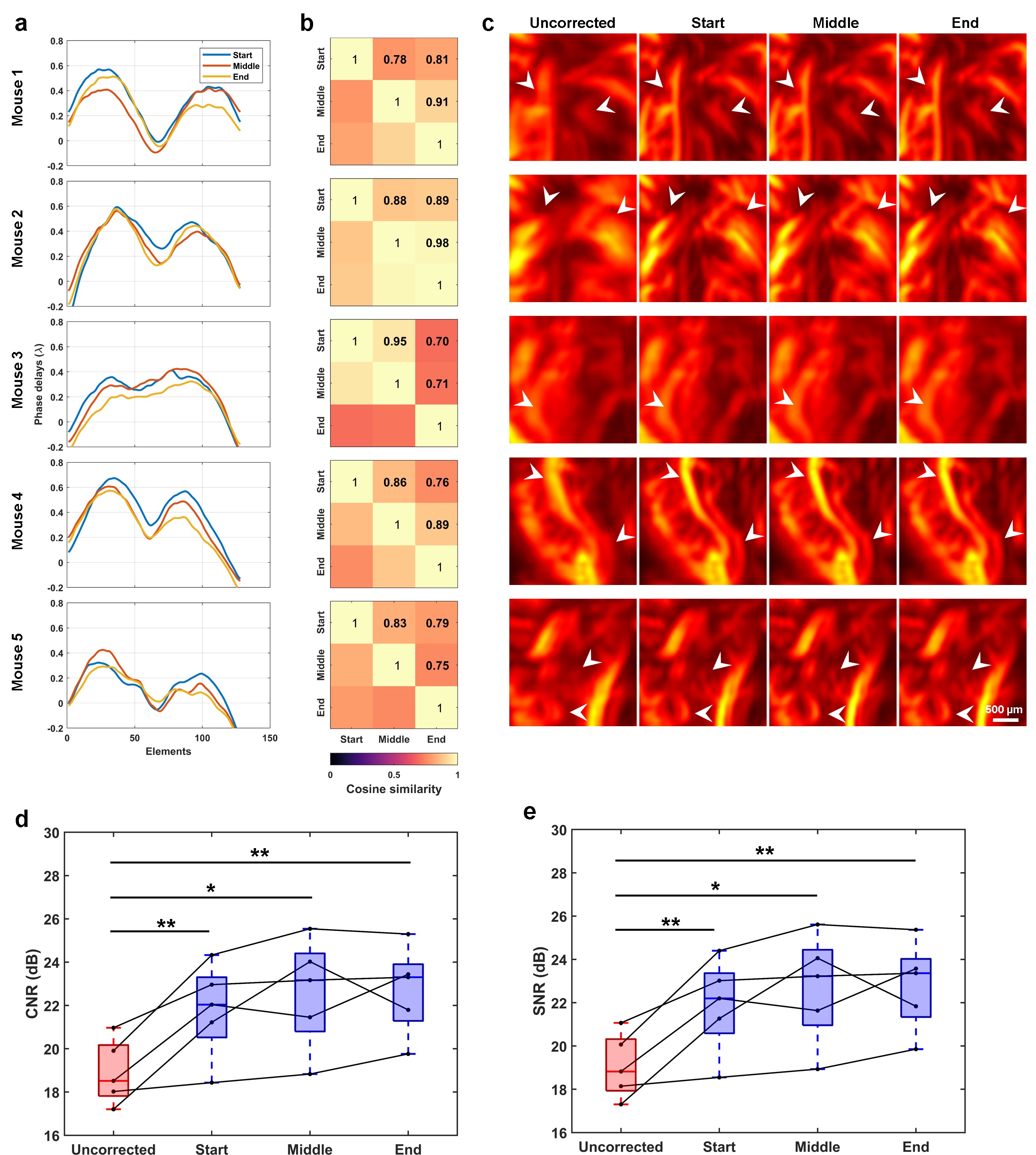}
    \caption{Impact of microbubble concentrations on the aberration function. (a) Aberration functions obtained at the start, middle, and end of the bolus emulating high, average, and low microbubble concentrations. (b) Modified cosine similarity of the aberration functions for the different concentrations. (c) Aberration correction on Power Doppler images for the aberration functions computed at different concentrations. (d) CNR  measured in CEUS Power Doppler images at different concentrations. No statistical difference was detected between concentrations (one-way ANOVA, p = 0.79) (e) SNR measured in CEUS Power Doppler images at different concentrations. No statistical difference was detected between concentrations (one-way ANOVA, p = 0.79 ). }
    \label{fig:concentration}
\end{figure}

\section{Results}

\subsection{Aberration correction on simulated images}

We first validated IPAC on simulated vascular data acquired with either a linear or a phased array probe. Fig. \ref{fig:in_silico}a and Fig. \ref{fig:in_silico_phased_array}a show the ground truth of the simulated vascular network. Increasing noise level in the RF channel data reduced the visibility of microbubble diffraction hyperbolas, as shown on  Fig. \ref{fig:in_silico}b and Fig. \ref{fig:in_silico_phased_array}b. Fig. \ref{fig:in_silico}c shows comparable aberration function profile to the ground truth was achieved after 6 iterations for the linear probe, while similar profile with a phase shift was observed for the phased array probe as shown on  Fig. \ref{fig:in_silico}c. The correction method remains robust even for SNR as low as 20 dB for both probe types, as shown on Fig. \ref{fig:in_silico}d and Fig. \ref{fig:in_silico_phased_array}d. Indeed, in presence of aberration, the vessel network is blurred while the corrected images showed sharper vessels.

\subsection{Aberration correction on \textit{In vivo} power Doppler images}

 An example of \textit{in vivo} microbubble signals used for aberration correction is shown in Fig. \ref{fig:invivo_microbubble}a, with detected microbubble positions marked in red. The corresponding \textit{in vivo} and simulated RF channel signals after SVD filtering and simulated RF channel signals using the microbubble positions are shown in Fig.  \ref{fig:invivo_microbubble}b. The microbubble positions were then used to construct the forward matrix shown in Fig. \ref{fig:invivo_microbubble}c. The correction method was able to improve microbubble shapes as shown in Fig. \ref{fig:invivo_microbubble}d. Close-ups show that microbubbles were more symmetrical and better focused after correction. Fig. \ref{fig:doppler}a shows the aberration function retrieved for different iteration steps on 5 mouse brains. The aberration functions reached a stable profile for all 5 mouse brains, with limited variation after 3–5 iterations, as shown in Fig. \ref{fig:doppler}d. The aberration profiles also showed a decrease in relative phase delays at the center of the probe for most Bregma sections identified. After correction, the CEUS Power Doppler signal is qualitatively improved in the whole-brain. Notably, sharper vessels can be observed, as indicated by arrows on Fig. \ref{fig:doppler}b. Close-up regions show blurred areas in presence of aberration appeared not only sharper but also in higher details after correction as shown on Fig. \ref{fig:doppler}c. 

Quantitatively, Fig. \ref{fig:doppler}e shows that the CNR was significantly improved from 18.9 dB to 23.5 dB (one-tailed paired t-test, p $<$ 0.0001), which represents an increase of 24$\%$. Similarly, the SNR increased by 23$\%$ from 19.1 dB to 23.6 dB after correction (one-tailed paired t-test, p $<$ 0.0001).

\begin{figure*}[ht!]
    \centering
    \includegraphics[width=1\linewidth]{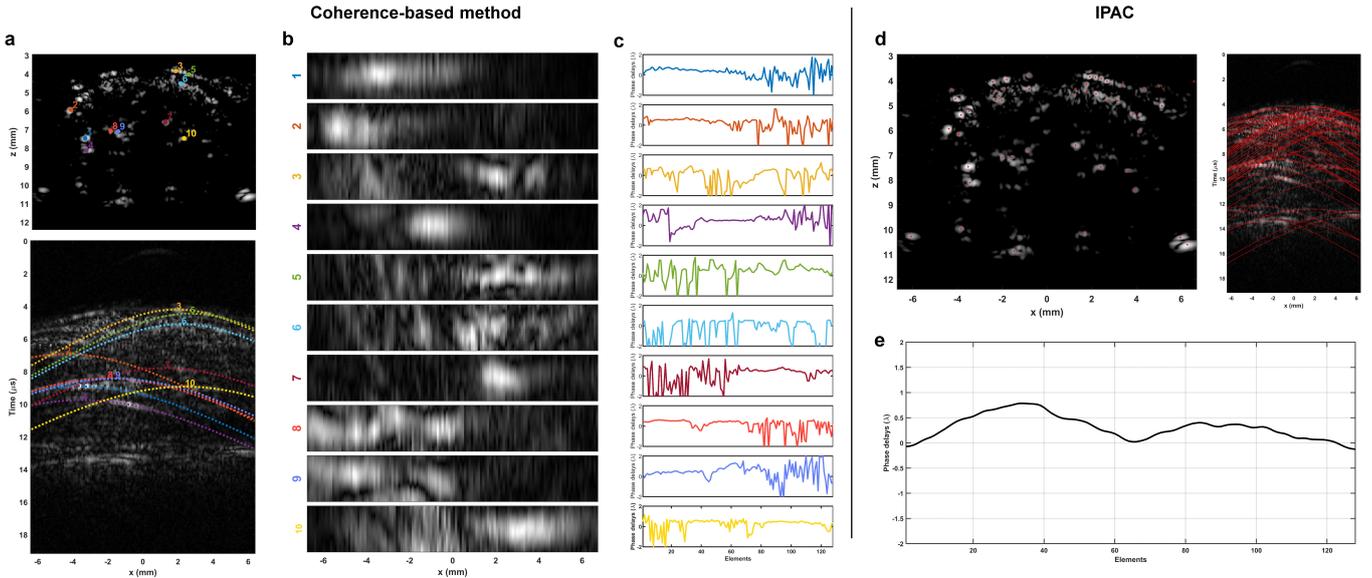}
    \caption{Comparison of IPAC performance on aberration functions computation with the coherence-based method. (a) Selected microbubbles with corresponding hyperbolas in the RF signals. (b) Realigned microbubble hyperbolas shown in (a). High directivity of ultrasound at 15 MHz limits microbubble signals to only a few elements. (c) Aberration profiles computed with the coherence-based method. Line colors correspond to identified microbubbles and hyperbolas shown in (a). Interference patterns and directivity of signals cause disruption in the computed aberration profiles. (d) Detected microbubbles with associated hyperbolas used in IPAC for a single frame. (e) Aberration function obtained with IPAC using a single frame and a single iteration.}
    \label{fig:benchmarking}
\end{figure*}

\begin{figure*}[ht!]
    \centering
    \includegraphics[width=1\linewidth]{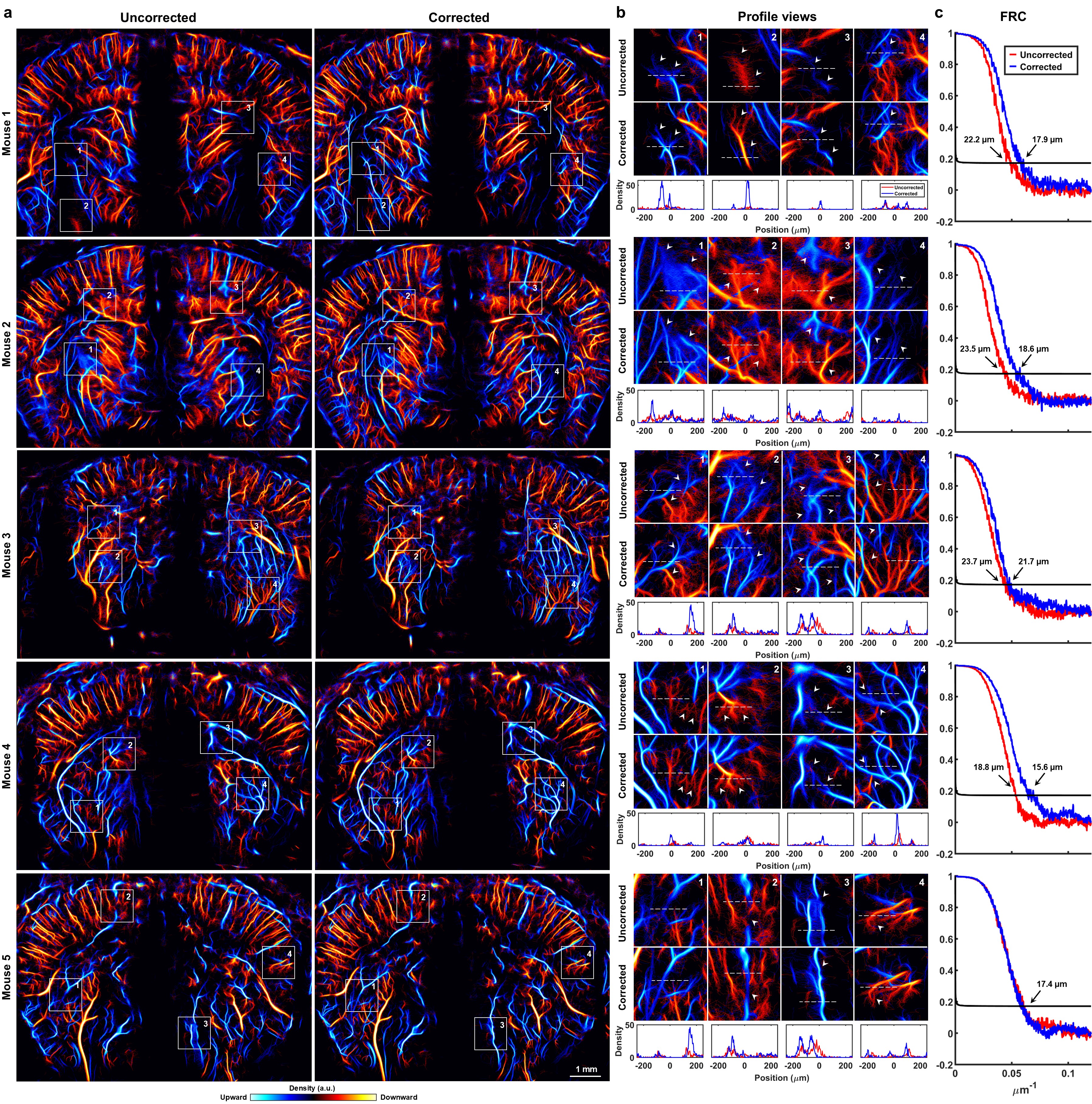}
    \caption{\textit{In vivo} aberration correction on transcranial ULM images of 5 different mouse brains. (a) Comparison of signed vascular maps before and after correction. (b) Regions of interest extracted from (a). Profile view comparisons of selected microvessels show an increase in resolution and in track densities after correction. (c) Resolution measurement with the FRC before and after correction. The resolution is improved for all mice except mouse 5.}
    \label{fig:invivo_ULM}
\end{figure*}

\begin{figure*}[ht!]
    \centering
    \includegraphics[width=1\linewidth]{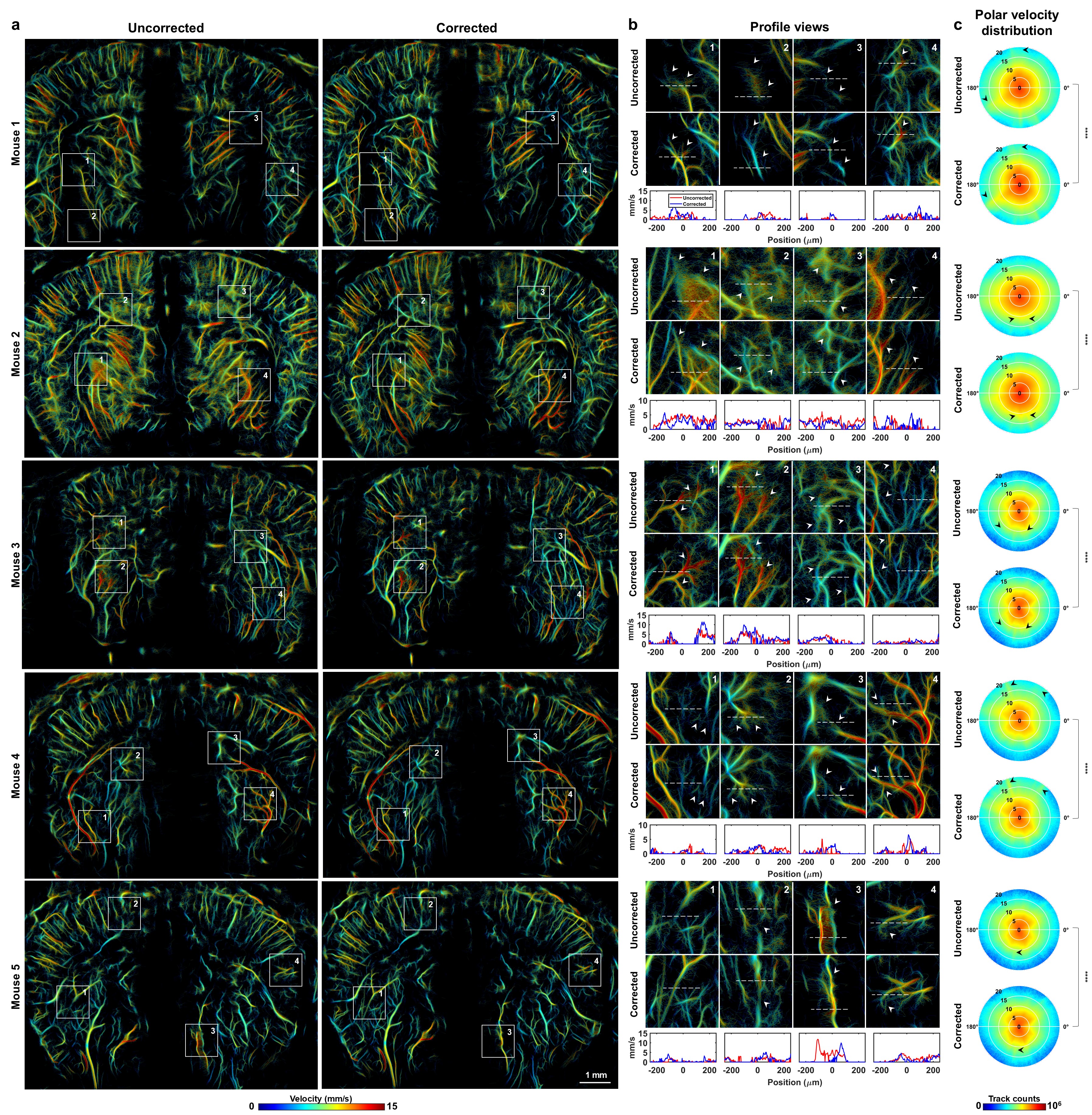}
    \caption{Evaluation of aberration correction on ULM hemodynamic quantification of 5 different mouse brains. (a) Comparison of velocity maps before and after correction. (b) Regions of interest extracted from (a). Selected microvessels show a more robust parabolic flow profile after correction. (c) Polar histogram showing the track velocity distribution. Flow direction before and after correction was statistically different for all mice (Watson-Williams test, p $<$ 0.0001). Radius represents velocity magnitude in mm/s and angle represents flow direction. After correction, the velocity distribution is more directive, as indicated by the black arrows.}
    \label{fig:invivo_ULM_vel}
\end{figure*}

\subsection{Microbubble concentration dependence of the aberration function}

The impact of microbubble concentration on aberration correction performance was evaluated by using buffers from the start, middle, and end of the bolus injection, which represent high, average, and low concentration conditions, respectively. The obtained aberration functions remained consistent with limited variation at different time of the injection for most mice, as shown in Fig. \ref{fig:concentration}a. Quantitatively, Fig. \ref{fig:concentration}b shows that the modified cosine similarity between the aberration functions obtained at different concentrations ranged between 0.70 and 0.98, with a mean value of 0.83. Fig. \ref{fig:concentration}c presents a comparison of the brain vasculature obtained after aberration correction using the different functions at different microbubble concentrations. The same close-up regions shown in Fig. \ref{fig:doppler} are presented. Improvement in vessels sharpness can be observed for all concentrations.

Fig.\ref{fig:concentration}d-e show that both CNR and SNR increased after correction for the different microbubble concentrations. No statistical difference was detected between the different concentrations for CNR (one-way ANOVA, p = 0.79) or for SNR (one-way ANOVA, p = 0.79). All aberration correction performed at the different concentrations significantly improved CNR and SNR (one-tailed paired t-test, p $<$ 0.05).

\subsection{Comparison to the coherence-based method}

In order to demonstrate the efficacy of IPAC at higher frequency and when multiple microbubbles are present, we compared it against the coherence-based method proposed in \cite{flax1988phase, demene2021transcranial, robin2023vivo}. Fig. \ref{fig:benchmarking}a shows examples of detected microbubbles with their corresponding hyperbolas in the channel data. Because of the high directivity of ultrasound at 15 MHz, rephased microbubble hyperbolas displayed limited signals to only a few dozen elements of the probe, as shown in Fig. \ref{fig:benchmarking}b. Interference patterns at hyperbolas crossing points can also be observed. Aberration functions computed with the coherence-based method show important discontinuities and high phase variations at interfering points and where elements had limited signals, as shown in Fig. \ref{fig:benchmarking}c.

Next, we evaluated IPAC performance on the same imaged frame used for the coherence-based method and for a single iteration. Fig. \ref{fig:benchmarking}d shows the detected microbubbles with their corresponding hyperbolas in the channel data used for the computation of IPAC. Fig. \ref{fig:benchmarking}e shows that the aberration function is continuous all along the elements, with a stable phase variation profile.

\subsection{Aberration correction on \textit{In vivo} ULM images}

Aberration correction also improved ULM reconstruction, as shown on Fig. \ref{fig:invivo_ULM}, which compares the signed vascular maps obtained before and after correction in five different mice. A higher number of microvessels can be observed after aberration correction, as well as a reduced number of blurred areas in multiple regions in the brain. Some duplicated vessels in presence of aberration (see mouse 2 and mouse 5) were properly connected after aberration correction. Selected close-ups and profile views presented in Fig. \ref{fig:invivo_ULM}b further demonstrates that contrast and resolution of microvessels are improved after correction. Additional or sharper vessels could be clearly identified after correction. Profile views showed more precisely that vessel widths are decreased, and contain a higher density of microbubbles. Quantitatively, FRC measurement showed that global ULM image resolution was improved for all mice, except for mouse 5. Indeed, as shown in Fig. \ref{fig:invivo_ULM}c for all other mice, the FRC curve moved to the right (i.e., in the direction of higher spatial frequency), which corresponds to a higher resolution.

Hemodynamic quantification of blood flow velocities is also improved after correction, as shown on Fig. \ref{fig:invivo_ULM_vel}. Fig. \ref{fig:invivo_ULM_vel}b shows the same close-up regions and selected vessel profiles previously identified on the anatomical map. Reconstructed vessels velocities are less spurious and show a more robust parabolic flow profile. After correction, track velocities are also more directive, as shown on the polar distributions of Fig. \ref{fig:invivo_ULM_vel}c, which encode both blood flow magnitude and direction. Indeed, sharper peaks can be identified in the polar plots, which represent the flow direction of blood vessels.
Statistical difference between flow direction before and after correction was confirmed with the Watson-Williams test for all mice (p $<$ 0.0001).

\begin{figure}[t]
    \centering
    \includegraphics[width=1\linewidth]{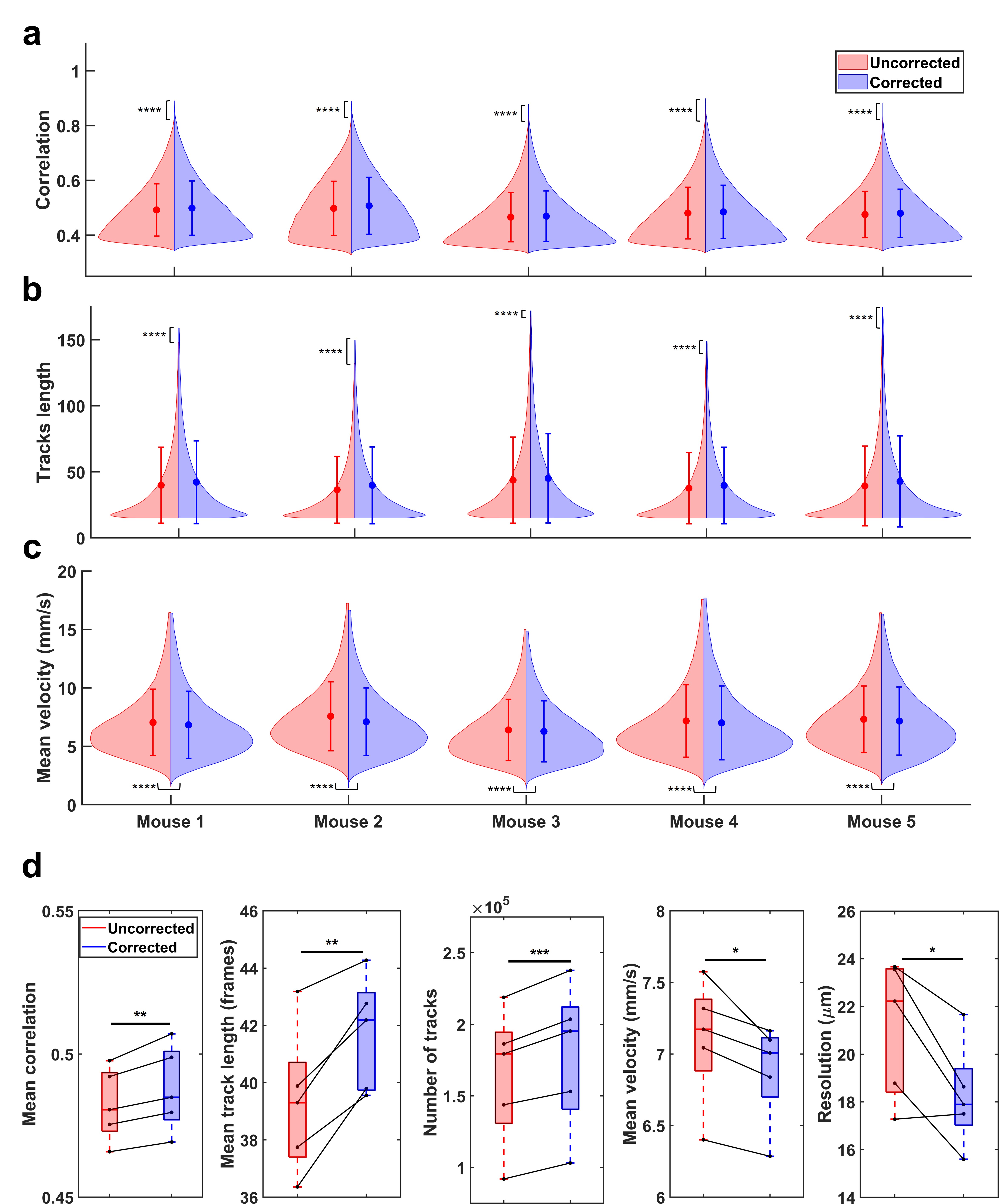}
    \caption{Quantitative analysis of difference before and after correction for ULM images. (a) Distribution of the microbubble correlation with the PSF before and after correction. The microbubble correlation is increased for all mice after correction (one-tailed KS test, p $<$ 0.0001). (b) Distribution of the track length before and after correction. The track length is increased for all mice after correction (one-tailed KS test, p $<$ 0.0001). (c) Distribution of the mean track velocities before and after correction.  The mean velocity is decreased for all mice after correction (one-tailed KS test, p $<$ 0.0001). (d) Assessment of the reproducibility of the correction performance on \textit{in vivo} ULM images across animals. The mean correlation of microbubbles with the PSF is increased after correction (one-tailed paired t-test, p $<$ 0.01). The number of tracks is increased after correction (one-tailed paired t-test, p $<$ 0.001). The mean track length is increased after correction (one-tailed paired t-test, p $<$ 0.01). The mean track velocity is decreased after correction (one-tailed paired t-test, p $<$ 0.05). The resolution in improved after correction (one-tailed paired t-test, p $<$ 0.05).}
    \label{fig:quantitative_each_mouse}
\end{figure}

Fig. \ref{fig:quantitative_each_mouse} shows a more detailed quantitative analysis of the correction performance in each mouse. Overall, the correction method moved the microbubble correlation and track length distributions to larger values for all mice (KS test, p $<$ 0.0001). We also observed a significant change in the distribution of mean velocities of track after correction (KS test, p $<$ 0.0001) with a higher number of tracks with lower velocities which are associated to smaller vessels as shown on Fig. \ref{fig:quantitative_each_mouse}c. Furthermore, Fig. \ref{fig:quantitative_each_mouse}d shows the reproducibility performance of the correction method when taking all animals into account. Our results showed that the mean microbubble correlation increased by 1.2 $\%$ (one-tailed t-test p $< 0.01$). On average, the mean track length increased by 6.2 $\%$ (one-tailed t-test, p $< 0.01$) and the total number of tracks increased by 8.8 $\%$ (one-tailed t-test, p $< 0.001$). The mean track velocity decreased from 7.10 mm/s to 6.88 mm/s (one-tailed t-test, p $< 0.05$). Finally, the resolution was improved by 13.5 $\%$ from 21.1 $\mu$m to 18.3 $\mu$m (one-tailed t-test, p $< 0.05$).

\section{Discussion}

In this work, we developed IPAC, an inverse problem approach to perform aberration correction by leveraging the sparsity of CEUS data and \textit{a pripori} knowledge of the medium based on microbubble localization. We introduced a gradient operator to express the forward problem in linear form with respect to the aberration, which allowed us to apply typical inversion strategies that minimize the mismatch between the data and the model to retrieve the aberration function. We first validated IPAC on simulated data using a linear and then a phased array probe to demonstrate that it can be generalized for plane wave as well for divergent wave imaging. We showed that IPAC can retrieve the proper aberration profile with respect to the ground truth. We also evaluated the reproducibility of transcranial aberration correction \textit{in vivo} in 5 mouse brains imaged at different coronal planes. By considering a single global aberration function, we were able to significantly improve image quality for both CEUS and ULM images. We showed that both CNR and SNR were improved on power Doppler images by $24\%$ and $23\%$, respectively. We also showed that IPAC performance was robust to microbubble concentrations and outperformed the coherence-based method when crossing hyperbolas are present and when the ultrasound directivity is high.

For ULM images, IPAC improved resolution by $13.5\%$, increased the mean tracks length by $6.2\%$ and the number of detected tracks by $8.8\%$. We also investigated the effect of aberration on velocity measurements and showed that both flow magnitude and direction could be misrepresented, as shown by other studies on blood flow imaging \cite{ivancevich2006phase,lindsey20143, fiorentini2023analysis}. Hence, aberration correction could be critical for establishing reliable biomarkers based upon hemodynamic variation. However, further investigations in presence of ground truths are required to further validate the impacts of aberration on blood velocities in ULM.

Recently proposed matrix formalism  taking advantage of angular coherence\cite{bendjador2020svd} or local spatial correlation \cite{lambert2020reflection} of ultrasound images have been used for aberration correction. We aimed to address a correction method for transcranial brain hemodynamic imaging. Imaging modalities such as ultrafast Doppler and ULM require being able to perform correction on hundreds or thousands of images for a single acquisition while maintaining a high frame rate. The advantage of our matrix formalism is that it can retrieve the aberration function from raw channel data without any change in the imaging sequence and then include the aberration correction into the beamforming.


Interestingly, the \textit{in vivo} aberration profiles we observed showed a decrease in phase delays at the center of the probe for most imaged planes. This is actually consistent with the variation of the skull thickness in the mouse, which is thicker at the middle \cite{liang2019impacts}. However, phase delays also depend on the propagation velocity in the forward model and using a different speed of sound could have resulted in a different aberration profile.  Our results showed a more significant improvement in Doppler images than in ULM images. This could be explained by the fact that tracking and smoothing algorithms can partially alleviate detection error or deformation caused by the skull aberration.

Moreover, figure \ref{fig:invivo_ULM} shows that some duplicated or disjoint vessels in presence of aberration are properly connected after correction (see mouse 2 and mouse 5). Such artifacts have been previously reported \cite{robin2023vivo, xing2023phase}. This could be mainly caused by aberrated microbubbles that are so distorted to the point they appear as two separated microbubbles, leading to serious localization errors or double localization. Aberration correction helped to restore the microbubble to their original shape and therefore removed duplicated vessels. Although aberration correction in mouse 5 improved some vessel appearance, the resolution did not improve. Since the FRC measures the image resolution in is globality \cite{hingot2021measuring}, it is possible that small local improvement did not translate into a global change in resolution.

In a broader sense, we presented an aberration correction formalism which general approach consists of defining a direct operator linking an aberration model (i.e., phase screen) to the ultrasound signals, combined with an optimization scheme (i.e. least-square inversion) to retrieve the aberration function. Such representation was possible by taking advantage of a matrix forward wave propagation model and by formulating the analytical expression of the gradient operator with respect to the aberration function. This formalism not only streamlines the process of aberration correction, it also takes into account a larger amount of information by including additional a priori knowledge of the medium, but also of the transmit pulse and probe characteristics. Another strength of using such a representation of aberration correction is that it could in principle be extended to other aberration models (e.g., localized or speed of sound map) or other optimization strategies (e.g., matched filter, gradient ascent/descent).

Although this work demonstrated that inverse problems can be used to perform phase aberration correction, some limitations remain. The forward model we proposed take into account only a unique phase screen aberration function. The phase screen model is a relatively simple model that may not fully capture the aberrating effects of the skull. Indeed, For biological tissues, the aberration function does not always remain consistent in the entire field of view \cite{dahl2005spatial}. This was also recently reported for transcranial imaging\cite{demene2021transcranial, robin2023vivo,xing2023phase}. Interestingly, for transcranial imaging using a phased array probe in humans, aberration profiles seem to have strong sectorial variation but limited depth-dependence \cite{robin2023vivo}. Still, IPAC could be adapted for depth dependent aberration correction by applying the inversion for different time windows through the RF channel data which correspond to different depths in the image. However, further examination on the number of time samples and/or number of frequencies required for inversion stability is required. 

Similarly, lateral variations could have been taken into account by using the DORT method \cite{prada1996decomposition} to isolate hyperbolas from different lateral regions, allowing to apply IPAC to recover a specific aberration function for each region. A more complete aberration function model with pixel-wise spatial variations to consider the skull thickness could also have been included into our formalism but would have required higher order derivatives of tensors. The use of open-source libraries such as Google JAX or PyTorch that can perform automatic differentiation could also facilitate implementation of IPAC without the explicit need of the analytical expression of the gradient operator.

Correction was applied only in receive using standard DAS beamforming. Other beamforming algorithms such as SMIF\cite{berthon2018spatiotemporal} or the SVD beamformer\cite{bendjador2020svd} could have been used to apply correction also in transmit. We considered the same aberration function for both transmit and receive, a valid assumption for the near-field phase screen model. However, more complex models may require considering different functions. A different aberration in transmit and receive could also have been included into our formalism but would have required to formulated two optimization problems to retrieve both aberration functions. We also presented our formalism for 2D imaging, the approach is general enough to be extended to 3D imaging since vectorization of both RF channel signals and aberration functions can be used to reduce the dimensionality of the problem. However, since 3D imaging can require thousands of elements, this raises the complexity of the computational need to solve the inverse problem. 

Although microbubbles can be considered as point scatterers for imaging, variations in properties such as microbubble sizes could affect the response spectrum. We aimed to partially address this issue by considering the backscattering amplitude of the microbubbles while constructing the forward model. However, further development may be required to better integrate microbubble properties into the inverse problem to ensure convergence of the solution.

The developed method relies on the presence of guided stars and prior knowledge of the medium. Hence, the ability of IPAC to properly retrieve the aberration function depends on the reliability of the localization algorithm, since false or missed detections could induce error in the estimation of the model. We previously introduced correlation with the PSF\cite{milecki2021deep, bourquin2021vivo} in order to improve detection rates. Finally, it would be also interesting to develop a similar optimization-based approach that could be used in a blind manner (i.e., without prior knowledge of the image) as other recently proposed methods\cite{van2023ultrasonic}, although their applications for \textit{in vivo} transcranial imaging still need to be demonstrated.

\section{Conclusion}
We developed IPAC, an inverse problem approach to aberration correction that directly links the aberrated ultrasound RF channel signals to the aberration functions by using a sparse representation of microbubble signals. For trancranial imaging of the mouse brain, we showed that IPAC improved Power Doppler contrast as well as ULM resolution. Finally, IPAC could be a promising aberration correction method for more reliable transcranial imaging of the brain vasculature with potential non-invasive clinical applications.

\bibliographystyle{ieeetran}
\bibliography{references.bib}

\begin{thebibliography}{10}
\providecommand{\url}[1]{#1}
\csname url@samestyle\endcsname
\providecommand{\newblock}{\relax}
\providecommand{\bibinfo}[2]{#2}
\providecommand{\BIBentrySTDinterwordspacing}{\spaceskip=0pt\relax}
\providecommand{\BIBentryALTinterwordstretchfactor}{4}
\providecommand{\BIBentryALTinterwordspacing}{\spaceskip=\fontdimen2\font plus
\BIBentryALTinterwordstretchfactor\fontdimen3\font minus
  \fontdimen4\font\relax}
\providecommand{\BIBforeignlanguage}[2]{{%
\expandafter\ifx\csname l@#1\endcsname\relax
\typeout{** WARNING: IEEEtran.bst: No hyphenation pattern has been}%
\typeout{** loaded for the language `#1'. Using the pattern for}%
\typeout{** the default language instead.}%
\else
\language=\csname l@#1\endcsname
\fi
#2}}
\providecommand{\BIBdecl}{\relax}
\BIBdecl

\bibitem{iadecola2017neurovascular}
C.~Iadecola, ``The neurovascular unit coming of age: a journey through
  neurovascular coupling in health and disease,'' \emph{Neuron}, vol.~96,
  no.~1, pp. 17--42, 2017.

\bibitem{montaldo2009coherent}
G.~Montaldo, M.~Tanter, J.~Bercoff, N.~Benech, and M.~Fink, ``Coherent
  plane-wave compounding for very high frame rate ultrasonography and transient
  elastography,'' \emph{IEEE transactions on ultrasonics, ferroelectrics, and
  frequency control}, vol.~56, no.~3, pp. 489--506, 2009.

\bibitem{tanter2014ultrafast}
M.~Tanter and M.~Fink, ``Ultrafast imaging in biomedical ultrasound,''
  \emph{IEEE transactions on ultrasonics, ferroelectrics, and frequency
  control}, vol.~61, no.~1, pp. 102--119, 2014.

\bibitem{mace2011functional}
E.~Mac{\'e}, G.~Montaldo, I.~Cohen, M.~Baulac, M.~Fink, and M.~Tanter,
  ``Functional ultrasound imaging of the brain,'' \emph{Nature methods},
  vol.~8, no.~8, pp. 662--664, 2011.

\bibitem{christensen2014vivo}
K.~Christensen-Jeffries, R.~J. Browning, M.-X. Tang, C.~Dunsby, and R.~J.
  Eckersley, ``In vivo acoustic super-resolution and super-resolved velocity
  mapping using microbubbles,'' \emph{IEEE transactions on medical imaging},
  vol.~34, no.~2, pp. 433--440, 2014.

\bibitem{errico2015ultrafast}
C.~Errico, J.~Pierre, S.~Pezet, Y.~Desailly, Z.~Lenkei, O.~Couture, and
  M.~Tanter, ``Ultrafast ultrasound localization microscopy for deep
  super-resolution vascular imaging,'' \emph{Nature}, vol. 527, no. 7579, pp.
  499--502, 2015.

\bibitem{bourquin2021vivo}
C.~Bourquin, J.~Por{\'e}e, F.~Lesage, and J.~Provost, ``In vivo pulsatility
  measurement of cerebral microcirculation in rodents using dynamic ultrasound
  localization microscopy,'' \emph{IEEE Transactions on Medical Imaging},
  vol.~41, no.~4, pp. 782--792, 2021.

\bibitem{perrot2021so}
V.~Perrot, M.~Polichetti, F.~Varray, and D.~Garcia, ``So you think you can das?
  a viewpoint on delay-and-sum beamforming,'' \emph{Ultrasonics}, vol. 111, p.
  106309, 2021.

\bibitem{anderson2000impact}
M.~Anderson, M.~McKeag, and G.~Trahey, ``The impact of sound speed errors on
  medical ultrasound imaging,'' \emph{The Journal of the Acoustical Society of
  America}, vol. 107, no.~6, pp. 3540--3548, 2000.

\bibitem{pinton2011sources}
G.~F. Pinton, G.~E. Trahey, and J.~J. Dahl, ``Sources of image degradation in
  fundamental and harmonic ultrasound imaging using nonlinear, full-wave
  simulations,'' \emph{IEEE transactions on ultrasonics, ferroelectrics, and
  frequency control}, vol.~58, no.~4, pp. 754--765, 2011.

\bibitem{mccall2021characterization}
J.~R. McCall, P.~A. Dayton, and G.~F. Pinton, ``Characterization of the
  ultrasound localization microscopy resolution limit in the presence of image
  degradation,'' \emph{IEEE Transactions on Ultrasonics, Ferroelectrics, and
  Frequency Control}, 2021.

\bibitem{demene2021transcranial}
C.~Demen{\'e}, J.~Robin, A.~Dizeux, B.~Heiles, M.~Pernot, M.~Tanter, and
  F.~Perren, ``Transcranial ultrafast ultrasound localization microscopy of
  brain vasculature in patients,'' \emph{Nature Biomedical Engineering},
  vol.~5, no.~3, pp. 219--228, 2021.

\bibitem{robin2023vivo}
J.~Robin, C.~Demen{\'e}, B.~Heiles, V.~Blanvillain, L.~Puke, F.~Perren, and
  M.~Tanter, ``In vivo adaptive focusing for clinical contrast-enhanced
  transcranial ultrasound imaging in human,'' \emph{Physics in Medicine \&
  Biology}, vol.~68, no.~2, p. 025019, 2023.

\bibitem{xing2023phase}
P.~Xing, J.~Por{\'e}e, B.~Rauby, A.~Malescot, E.~Martineau, V.~Perrot, R.~L.
  Rungta, and J.~Provost, ``Phase aberration correction for in vivo ultrasound
  localization microscopy using a spatiotemporal complex-valued neural
  network,'' \emph{IEEE Transactions on Medical Imaging}, vol.~43, no.~2, pp.
  662--673, 2023.

\bibitem{mace2013functional}
E.~Mace, G.~Montaldo, B.-F. Osmanski, I.~Cohen, M.~Fink, and M.~Tanter,
  ``Functional ultrasound imaging of the brain: theory and basic principles,''
  \emph{IEEE transactions on ultrasonics, ferroelectrics, and frequency
  control}, vol.~60, no.~3, pp. 492--506, 2013.

\bibitem{o1988phase}
M.~O'donnell and S.~Flax, ``Phase-aberration correction using signals from
  point reflectors and diffuse scatterers: Measurements,'' \emph{IEEE
  transactions on ultrasonics, ferroelectrics, and frequency control}, vol.~35,
  no.~6, pp. 768--774, 1988.

\bibitem{flax1988phase}
S.~Flax and M.~O'Donnell, ``Phase-aberration correction using signals from
  point reflectors and diffuse scatterers: Basic principles,'' \emph{IEEE
  transactions on ultrasonics, ferroelectrics, and frequency control}, vol.~35,
  no.~6, pp. 758--767, 1988.

\bibitem{maasoy2005iteration}
S.-E. M{\aa}s{\o}y, T.~Varslot, and B.~Angelsen, ``Iteration of transmit-beam
  aberration correction in medical ultrasound imaging,'' \emph{The Journal of
  the Acoustical Society of America}, vol. 117, no.~1, pp. 450--461, 2005.

\bibitem{montaldo2011time}
G.~Montaldo, M.~Tanter, and M.~Fink, ``Time reversal of speckle noise,''
  \emph{Physical review letters}, vol. 106, no.~5, p. 054301, 2011.

\bibitem{osmanski2012aberration}
B.-F. Osmanski, G.~Montaldo, M.~Tanter, and M.~Fink, ``Aberration correction by
  time reversal of moving speckle noise,'' \emph{IEEE transactions on
  ultrasonics, ferroelectrics, and frequency control}, vol.~59, no.~7, pp.
  1575--1583, 2012.

\bibitem{ivancevich2008real}
N.~M. Ivancevich, G.~F. Pinton, H.~A. Nicoletto, E.~Bennett, D.~T. Laskowitz,
  and S.~W. Smith, ``Real-time 3-d contrast-enhanced transcranial ultrasound
  and aberration correction,'' \emph{Ultrasound in medicine \& biology},
  vol.~34, no.~9, pp. 1387--1395, 2008.

\bibitem{lindsey20143}
B.~D. Lindsey, H.~A. Nicoletto, E.~R. Bennett, D.~T. Laskowitz, and S.~W.
  Smith, ``3-d transcranial ultrasound imaging with bilateral phase aberration
  correction of multiple isoplanatic patches: A pilot human study with
  microbubble contrast enhancement,'' \emph{Ultrasound in medicine \& biology},
  vol.~40, no.~1, pp. 90--101, 2014.

\bibitem{soulioti2019super}
D.~E. Soulioti, D.~Esp{\'\i}ndola, P.~A. Dayton, and G.~F. Pinton,
  ``Super-resolution imaging through the human skull,'' \emph{IEEE transactions
  on ultrasonics, ferroelectrics, and frequency control}, vol.~67, no.~1, pp.
  25--36, 2019.

\bibitem{schoen2019heterogeneous}
S.~Schoen and C.~D. Arvanitis, ``Heterogeneous angular spectrum method for
  trans-skull imaging and focusing,'' \emph{IEEE transactions on medical
  imaging}, vol.~39, no.~5, pp. 1605--1614, 2019.

\bibitem{tian2023transcranial}
Z.~Tian, M.~Olmstead, Y.~Jing, and A.~Han, ``Transcranial phase correction
  using pulse-echo ultrasound and deep learning: a 2-d numerical study,''
  \emph{IEEE transactions on ultrasonics, ferroelectrics, and frequency
  control}, vol.~71, no.~1, pp. 117--126, 2023.

\bibitem{oreilly2013super}
M.~A. O'Reilly and K.~Hynynen, ``A super-resolution ultrasound method for brain
  vascular mapping,'' \emph{Medical physics}, vol.~40, no.~11, p. 110701, 2013.

\bibitem{lambert2020reflection}
W.~Lambert, L.~A. Cobus, M.~Couade, M.~Fink, and A.~Aubry, ``Reflection matrix
  approach for quantitative imaging of scattering media,'' \emph{Physical
  Review X}, vol.~10, no.~2, p. 021048, 2020.

\bibitem{garcia2022simus}
D.~Garcia, ``Simus: An open-source simulator for medical ultrasound imaging.
  part i: Theory \& examples,'' \emph{Computer Methods and Programs in
  Biomedicine}, p. 106726, 2022.

\bibitem{tanter2000time}
M.~Tanter, J.-L. Thomas, and M.~Fink, ``Time reversal and the inverse filter,''
  \emph{The Journal of the Acoustical Society of America}, vol. 108, no.~1, pp.
  223--234, 2000.

\bibitem{tanter2001optimal}
J.-F. Aubry, J.~Gerber, J.-L. Thomas, and M.~Fink, ``Optimal focusing by
  spatio-temporal inverse filter. i. basic principles,'' \emph{The Journal of
  the Acoustical Society of America}, vol. 110, no.~1, pp. 37--47, 2001.

\bibitem{tihonov1963solution}
A.~N. Tihonov, ``Solution of incorrectly formulated problems and the
  regularization method,'' \emph{Soviet Math.}, vol.~4, pp. 1035--1038, 1963.

\bibitem{tiran2015multiplane}
E.~Tiran, T.~Deffieux, M.~Correia, D.~Maresca, B.-F. Osmanski, L.-A. Sieu,
  A.~Bergel, I.~Cohen, M.~Pernot, and M.~Tanter, ``Multiplane wave imaging
  increases signal-to-noise ratio in ultrafast ultrasound imaging,''
  \emph{Physics in Medicine \& Biology}, vol.~60, no.~21, p. 8549, 2015.

\bibitem{milecki2021deep}
L.~Milecki, J.~Por{\'e}e, H.~Belgharbi, C.~Bourquin, R.~Damseh,
  P.~Delafontaine-Martel, F.~Lesage, M.~Gasse, and J.~Provost, ``A deep
  learning framework for spatiotemporal ultrasound localization microscopy,''
  \emph{IEEE Transactions on Medical Imaging}, vol.~40, no.~5, pp. 1428--1437,
  2021.

\bibitem{belgharbi2023anatomically}
H.~Belgharbi, J.~Por{\'e}e, R.~Damseh, V.~Perrot, L.~Milecki,
  P.~Delafontaine-Martel, F.~Lesage, and J.~Provost, ``An anatomically
  realistic simulation framework for 3d ultrasound localization microscopy,''
  \emph{IEEE Open Journal of Ultrasonics, Ferroelectrics, and Frequency
  Control}, vol.~3, pp. 1--13, 2023.

\bibitem{heiles2022performance}
B.~Heiles, A.~Chavignon, V.~Hingot, P.~Lopez, E.~Teston, and O.~Couture,
  ``Performance benchmarking of microbubble-localization algorithms for
  ultrasound localization microscopy,'' \emph{Nature Biomedical Engineering},
  pp. 1--12, 2022.

\bibitem{demene2015spatiotemporal}
C.~Demen{\'e}, T.~Deffieux, M.~Pernot, B.-F. Osmanski, V.~Biran, J.-L.
  Gennisson, L.-A. Sieu, A.~Bergel, S.~Franqui, J.-M. Correas \emph{et~al.},
  ``Spatiotemporal clutter filtering of ultrafast ultrasound data highly
  increases doppler and fultrasound sensitivity,'' \emph{IEEE transactions on
  medical imaging}, vol.~34, no.~11, pp. 2271--2285, 2015.

\bibitem{yan2023fast}
J.~Yan, B.~Wang, K.~Riemer, J.~Hansen-Shearer, M.~Lerendegui, M.~Toulemonde,
  C.~J. Rowlands, P.~D. Weinberg, and M.-X. Tang, ``Fast 3d super-resolution
  ultrasound with adaptive weight-based beamforming,'' \emph{IEEE Transactions
  on Biomedical Engineering}, vol.~70, no.~9, pp. 2752--2761, 2023.

\bibitem{zhao2018robust}
Z.~Zhao, H.~Zhang, Z.~Xiao, H.~Du, Y.~Zhuang, C.~Fan, and H.~Zhao, ``Robust 2d
  phase unwrapping algorithm based on the transport of intensity equation,''
  \emph{Measurement Science and Technology}, vol.~30, no.~1, p. 015201, 2018.

\bibitem{garcia2010robust}
D.~Garcia, ``Robust smoothing of gridded data in one and higher dimensions with
  missing values,'' \emph{Computational statistics \& data analysis}, vol.~54,
  no.~4, pp. 1167--1178, 2010.

\bibitem{akima1970new}
H.~Akima, ``A new method of interpolation and smooth curve fitting based on
  local procedures,'' \emph{Journal of the ACM (JACM)}, vol.~17, no.~4, pp.
  589--602, 1970.

\bibitem{hingot2021measuring}
V.~Hingot, A.~Chavignon, B.~Heiles, and O.~Couture, ``Measuring image
  resolution in ultrasound localization microscopy,'' \emph{IEEE transactions
  on medical imaging}, vol.~40, no.~12, pp. 3812--3819, 2021.

\bibitem{ivancevich2006phase}
N.~M. Ivancevich, J.~J. Dahl, G.~E. Trahey, and S.~W. Smith, ``Phase-aberration
  correction with a 3-d ultrasound scanner: Feasibility study,'' \emph{ieee
  transactions on ultrasonics, ferroelectrics, and frequency control}, vol.~53,
  no.~8, pp. 1432--1439, 2006.

\bibitem{fiorentini2023analysis}
S.~Fiorentini, S.-E. M{\aa}s{\o}y, and J.~Avdal, ``Analysis of aberration
  effects on flow imaging and quantification in echocardiography,'' \emph{IEEE
  Open Journal of Ultrasonics, Ferroelectrics, and Frequency Control}, vol.~3,
  pp. 194--202, 2023.

\bibitem{bendjador2020svd}
H.~Bendjador, T.~Deffieux, and M.~Tanter, ``The svd beamformer: Physical
  principles and application to ultrafast adaptive ultrasound,'' \emph{IEEE
  transactions on medical imaging}, vol.~39, no.~10, pp. 3100--3112, 2020.

\bibitem{liang2019impacts}
B.~Liang, W.~Liu, Q.~Zhan, M.~Li, M.~Zhuang, Q.~H. Liu, and J.~Yao, ``Impacts
  of the murine skull on high-frequency transcranial photoacoustic brain
  imaging,'' \emph{Journal of biophotonics}, vol.~12, no.~7, p. e201800466,
  2019.

\bibitem{dahl2005spatial}
J.~J. Dahl, M.~S. Soo, and G.~E. Trahey, ``Spatial and temporal aberrator
  stability for real-time adaptive imaging,'' \emph{IEEE transactions on
  ultrasonics, ferroelectrics, and frequency control}, vol.~52, no.~9, pp.
  1504--1517, 2005.

\bibitem{prada1996decomposition}
C.~Prada, S.~Manneville, D.~Spoliansky, and M.~Fink, ``Decomposition of the
  time reversal operator: Detection and selective focusing on two scatterers,''
  \emph{The Journal of the Acoustical Society of America}, vol.~99, no.~4, pp.
  2067--2076, 1996.

\bibitem{berthon2018spatiotemporal}
B.~Berthon, P.~Morichau-Beauchant, J.~Por{\'e}e, A.~Garofalakis, B.~Tavitian,
  M.~Tanter, and J.~Provost, ``Spatiotemporal matrix image formation for
  programmable ultrasound scanners,'' \emph{Physics in Medicine \& Biology},
  vol.~63, no.~3, p. 03NT03, 2018.

\bibitem{van2023ultrasonic}
P.~van~der Meulen, M.~Couti{\~n}o, J.~G. Bosch, P.~Kruizinga, and G.~Leus,
  ``Ultrasonic imaging through aberrating layers using covariance matching,''
  \emph{IEEE Transactions on Computational Imaging}, vol.~9, pp. 745--759,
  2023.

\end{thebibliography}

\section*{Supplemental materials}\label{section:appendix}

\subsection{Theoretical derivation of the transmission and reflection operators}

In this section, we describe in more details the theoretical derivation of the  transmission and reflection operators based upon the physics of wave propagation. First, the propagation of a monochromatic wave within a homogeneous medium is defined by the wave equation
\begin{align}
    \left(\nabla^2-\frac{1}{c^2}\frac{\partial}{\partial t^2}\right)\psi(\mathbf{r},t)=s(\mathbf{r},t),
\end{align}

with $c$ the medium wave velocity and $s(\mathbf{r},t)$ the source term. By considering a time dependence of the form $\psi(\mathbf{r},t)=\psi(\mathbf{r})e^{-i\omega t}$ and taking the Fourier transform, we can determine the wave propagation through the medium by finding solutions for the  Helmholtz equation
\begin{align}
    \left(\nabla^2+k^2 \right)\psi(\mathbf{r})=s(\mathbf{r}),
\end{align}
with $k=\omega/c$ the wavenumber. Considering each element as a point source, meaning that $s(\mathbf{r}-\mathbf{r}_n)=\delta(\mathbf{r}-\mathbf{r}_n)$, the wave propagation between each element at $\mathbf{r}_n$ to any point $\mathbf{r}_s$ in the medium is defined by the Green's function for the Helmholtz equation
\begin{align}
G_{n,s}(\omega)=G(\mathbf{r}_n,\mathbf{r}_s,\omega)=\frac{\exp\left({ik||\mathbf{r}_s-\mathbf{r}_n||}\right)}{4\pi ||\mathbf{r}_s-\mathbf{r}_n||}.
\end{align}

To properly emulate the transmitted field, additional physical aspects of an ultrasonic probe must be taken into consideration 
\cite{garcia2022simus}. The element transducer response $W_n(\omega) =W(\mathbf{r}_n,\omega)$, directivity $ D_{n,s}(\omega)=  D(\mathbf{r}_n,\mathbf{r}_s,\omega)$ and elevation focusing $ F_{n,s}(\omega)=F(\mathbf{r}_n,\mathbf{r}_s,\omega)$ will affect the spectral bandwidth as well the spatial distribution of the transmitted pulse. The element response $W$ represents the bandwidth of the transducer, generally defined with a central frequency $\omega_0$ and bandwidth $\Delta \omega$. The directivity take into account that each element does not emit or transmit uniformly in all direction, which will affect the energy distribution surrounding the element. The elevation focusing $F$ allows considering the element size in 2 dimensions. The emission by an element for a given pulse  $P_0(\omega)$ is then defined by
\begin{align}
 {T_0}_{n,s}(\omega)=  D_{n,s}(\omega)F_{n,s}(\omega) \sqrt{W_n(\omega)}P_0(\omega)e^{i\omega \Delta\tau_n},
\end{align}

where $\Delta\tau_n$ stands for the transmit delay associated to each element. This formalism allows us to define any kind of one-dimensional probe, for plane wave as well for divergent wave transmission. Taking into account the wave propagation, the complete transmitted field $T_{n,s}(\omega)$ is then given by
\begin{align}
T_{n,s}(\omega)&={T_0}_{n,s}(\omega)G_{n,s}(\omega).
\end{align}

Considering each scatterer at position $\mathbf{r}_s$ as point source, the reflection to the receiving element at position $\mathbf{r}_m$ is also given by the Green function $G_{m,s}(\omega)=G(\mathbf{r}_m,\mathbf{r}_s,\omega)$. The transducer properties will affect the received signal. Taking into consideration again the directivity, the elevation but also the transducer response, the modulation by the element in receive is given by
\begin{align}
{R_0}_{m,s}(\omega)=    \sqrt{W_m(\omega)}D_{m,s}(\omega)F_{m.s}(\omega).
\end{align}

The back propagation field is then given by
\begin{align}
R_{m,s}(\omega)&={R_0}_{m,s}(\omega)G_{m,s}(\omega).
\end{align}

\subsection{Additional definitions}

In this section we present additional details on the definition of the element directivity, elevation focusing and transducer response. The developments of the characteristics of the ultrasound element are adapted from \cite{garcia2022simus}.

\subsubsection{Directivity}
The model taken for simulating each element corresponds to a baffled piston vibrating in direction $z$. For an infinite baffle impedance, the directivity can be expressed as

\begin{align}
D&=\text{sinc}\big(\frac{kb}{2}\sin\alpha\big)
\end{align}

with $b$ the element width and $\alpha$ the angle between $\mathbf{r}=\mathbf{r}_s-\mathbf{r}_n$ and the $z$ axis.

\subsubsection{Elevation focusing}
The elevation focusing corresponds to the distance from the element where the pressure field is maximal and can be defined as 

\begin{align}
F=\int_{-h/2}^{h/2}e^{iky'^2/2r_f}e^{ik(y-y)'^2/2|\mathbf{r}|}dy',
\end{align}

with $h$ the element height and $r_f$ the distance at elevation focus.

The elevation focusing allows to consider the change in the pressure field depending of $|\mathbf{r}|$, the distance between element and the scatterer, and how focusing degrades according to the depth.

\subsubsection{Transducer response}

The spectrum of the transducer can be defined as a Gaussian window

\begin{align}
W &= \exp{\left[-\ln2\left(\frac{2|\omega-\omega_c|}{\omega_b}\right)^p\right]},
\end{align}

with $\omega_c$ the central frequency, $\omega_b$ the bandwidth, and

\begin{align}
    p=\ln126/\ln\left(2\frac{\omega_c}{\omega_b}\right).
\end{align}

We defined here $W$ as the two-way response of the transducer (in transmit and receive). Hence, the one-way response is given by $\sqrt{W}$.

\subsection{Regularization matrix}

The regularization matrix can be defined with the finite difference coefficients. The second derivative matrix using the centered difference is given by

\begin{align}
\mathbf{D_2}=
\begin{bmatrix}
-1&1&&&&\\
1&-2&1&&&\\
&\dots&\dots&\dots&&\\
&&1&-2&1&\\
&&&&-1&1
\end{bmatrix}.
\end{align}

\end{document}